\documentclass[prd,aps,floatfix,nofootinbib]{revtex4}

\usepackage{amsbsy}
\usepackage{amssymb}
\usepackage[sumlimits]{amsmath}
\usepackage{graphicx}
\usepackage[active]{srcltx}
\usepackage{pdfsync}
\usepackage{cleveref}
\usepackage{float}

\usepackage{color}

\begin{document}

\title{Thermal aspects of neutron star mergers}

\author{P. Hammond, I. Hawke and N. Andersson}

\affiliation{
Mathematical Sciences and STAG Research Centre, University of Southampton,
Southampton SO17 1BJ, United Kingdom}

\begin{abstract}
In order to extract maximal information from neutron-star merger signals, both gravitational and electromagnetic, we need to ensure that our theoretical models/numerical simulations faithfully represent the extreme physics involved. This involves a range of issues, with the finite temperature effects regulating many of the relevant phenomena. As a step towards understanding these issues, we explore the conditions for $\beta$-equilibrium in neutron star matter for the densities and temperatures reached in a binary neutron star merger. Using the results from our out-of-equilibrium merger simulation, we consider how different notions of equilibrium may affect the merger dynamics, raising issues that arise when attempting to account for these conditions in future simulations. These issues are both computational and conceptual. We show that the effects lead to, in our case,  a softening of the equation of state in some density regions, and to composition changes that affect processes that rely on deviation from equilibrium, such as bulk viscosity, both in terms of the magnitude and the equilibration timescales inherent to the relevant set of reactions. We also demonstrate that it is difficult to determine exactly which equilibrium conditions are relevant in which regions of the matter due to the dependence on neutrino absorption, further complicating the calculation of the reactions that work to restore the matter to equilibrium.
\end{abstract}

\maketitle

\section{Introduction}

In terms of cosmic fireworks, neutron star mergers rank (alongside supernovae) as the most spectacular. Reaching temperatures similar to those recorded in  intermediate-energy collider experiments, with matter compressed to a density that cannot be reached in the laboratory, these events involve physics at the limit of our understanding---and beyond. The excitement of observing such events in the full glory of multimessenger astronomy, including gravitational waves, is evident. The breakthrough observations of GW170817 \cite{2017ApJ...848L..12A,2017PhRvL.119p1101A,2019PhRvX...9a1001A}---including the late inspiral gravitational-wave signal, the short gamma-ray burst that followed the merger and the kilonova emission associated with nuclear reactions in the matter outflow \cite{2017ApJ...848L..17C}---raised the stakes by providing a wealth of information. However, the modelling of neutron star mergers still presents a severe challenge \cite{2017RPPh...80i6901B,2020ARNPS..70...95R}, which can only be met by large-scale numerical simulations with an increasing level of realism---conceptually difficult and computationally expensive. 

In this study, we focus our attention on the thermal aspects of the problem. This is important for a number of reasons. First of all, recent simulations have established that neutron star mergers may reach higher temperatures than one might naively expect (in some cases rising to almost 100~MeV \cite{2019EPJA...55..124P,Endrizzi2020,2021arXiv210607885P}) and it makes sense to ask to what extent this impacts on our understanding of the problem. Secondly, recent theoretical work points to thermal effects impacting on key physics aspects, like the meaning of chemical equilibrium \cite{Alford2018a,alford2021beta}. Again, we need to explore what impact this new understanding may have on the simulations and the interpretation of the results. This is a natural step towards higher level issues which aim to sharpen the questions we address with numerical relativity simulations. Which aspects of the rich physics are within reach and which are not? 

\section{The current state of the art}

\subsection{The equation of state}

While it may be tempting to focus the discussion on different aspects of our numerical simulations, the key points we aim to make relate to the thermodynamics and the microphysics encoded in the matter equation of state. This means that we need to keep a keen eye on the input physics---and to what extent this physics is encoded in the simulation. Given this, and to provide the proper context for the discussion, it makes sense to first take a closer look at the  assumptions associated with the simulated matter model.

Neutron stars are complex physical systems, involving a range of aspects from nuclear physics and condensed matter physics through to electromagnetism and relativistic gravity. As the different issues are (inevitably) linked, the assembly of even moderately realistic models is a challenge. Unless we want to get stuck at the very first step, we have to make simplifications. A natural starting point is to describe the star as multi-fluid system. There are two good reasons for this. First, we have a good intuitive understanding of (at least the basics of)  the relevant fluid dynamics. Second, the formalism required to model relativistic multi-fluid systems is fairly well developed \cite{2021LRR....24....3A}. It is then relatively easy to argue \cite{2017CQGra..34l5003A} that we need to keep track of (at least) four fluid degrees of freedom for a moderately cold neutron star. We have to consider the usual bulk flow along with the electromagnetic charge current, a heat flux and a relative drift of superfluid neutrons in the star's core. If we focus on neutron star mergers, the thermal aspects dominate and we inevitably have to consider the role of the neutrinos that are generated as the matter composition evolves. As we will see,  this complicates the problem in a number of ways. 

In view of these different issues, the assumptions required to reach the starting point for 
all current neutron star simulations may seem rather drastic. We ignore all the multi-fluid features and reduce the problem to a single (perfect) fluid model (or, in the case of magnetohydrodynamics, a single fluid alongside some simple closure assumption for the charge current). This retains aspects related to the matter composition and the heat, but ignores intricate details regarding the expected relative flows. A pragmatist might argue that this step is necessary to make progress---which may well be true---but a realist might want to urge some level of caution. We need to make sure that we are not trying to draw (too many) conclusions about physics that is not faithfully represented in the model.

The immediate advantage of the single-fluid model is that we can make a direct connection with the thermodynamics and the small scale physics. The natural starting point for this discussion is the first law of thermodynamics, which then states that 
the change in energy $dE$ of a many-particle-species system (with different species labelled by x, all moving together) is given by 
\begin{equation}
    dE = T dS - p dV + \sum_\text{x} \mu_\text{x} dN_\text{x},
    \label{eq:FirstLawFull}
\end{equation}
where $T$ is the temperature, $dS$ is the change in entropy, $p$ is the 
pressure, $dV$ is the change in volume and $\mu_\text{x}$ and $dN_\text{x}$ are 
the chemical potential (the energy associated with adding or removing a particle from the system)
and change in particle number, respectively, of  species $\text{x}$. The first 
term, $T dS$, may be identified as the heat put into the system; the second, 
$- p dV$, corresponds to the work done on the system; and the final term, 
$\mu_\text{x} dN_\text{x}$, is associated with a change in number of each particle 
species, e.g. due to nuclear reactions. Effectively, we have an equation of state for the system, 
$E(S, V, \{N_\text{x}\})$, but in practice it is more convenient to express the thermodynamics in terms of densities. This leads to a Gibbs relation (essentially the integrated and densitised version of \eqref{eq:FirstLawFull})
\begin{equation}
    e = sT - p + \sum_\text{x} n_\mathrm{x} \mu_\text{x} \ ,
    \label{eq:EulerRelation}
\end{equation}
where $e$ is the total energy density, $s$ is the 
entropy density and $n_\mathrm{x}$ is each species' number density. Noting that 
\begin{equation}
    T = \left( \frac{\partial e}{\partial s} 
        \right)_{\{n_\text{x}\}}, \quad  \mu_\text{x} = \left( \frac{\partial e}{\partial n_\text{x}} 
        \right)_{s,  \{n_\text{y}\}},
    \label{eq:Intensive1}
\end{equation}
with $\mathrm y \neq \mathrm x$, we see that the pressure is completely determined from an equation of state of form $e = e(n_\mathrm{x}, s)$. 
So far, the arguments are quite general.

The specific simulation we will consider implements the APR equation of state from \cite{Schneider2019} (as implemented in the CompOSE library \cite{CompOSE}). This means that we have matter composed of neutrons, protons and electrons~\footnote{For clarity, we ignore the muons even though they are expected to be present throughout the bulk of a neutron star's core. Including them would  not be very difficult as they enter the problem in the same way as the electrons.}   (n, p and e). Neutrinos, although key to many of the aspects we will consider later, are not included in the  equation of state prescription (and hence will not be considered in the arguments we sketch at this point). 
If the system retains local charge neutrality (as we will assume) we then have $n_\mathrm{p}=n_\mathrm{e}$ and if we  introduce the baryon number density $n_\mathrm{b}=n_\mathrm{n}+n_\mathrm{p}$ and  the electron fraction  $Y_\mathrm{e} = n_\mathrm{e}/n_\mathrm{b}$  we have the three-parameter model $e=e(n_\mathrm{b}, Y_{\mathrm e}, s)$ or, equivalently (after a Legendre transform), $f=f(n_\mathrm{b}, Y_{\mathrm e}, T)$ where $f=e-Ts$ is the free energy density. The required information is typically provided as an equation of state table.

As much of our discussion will focus on issues relating to chemical equilibrium, let us explore what this involves. In general, the system is in chemical equilibrium with respect to the relevant
 reactions  when the  chemical potentials balance (we will discuss the details later in \cref{Sec:Results}). This typically leads to an equation which can be solved for the equilibrium matter composition, which then reduces the number of parameters by one. We may, for example, use the equilibrium relation to determine the electron fraction. In essence, we are then assuming that nuclear reactions act to reinstate equilibrium on a timescale much faster than the dynamics we intend to study \cite{2019MNRAS.489.4043A}. The upshot is that the equilibrium equation of state for our system has only two parameters,  $e=e(n_\mathrm{b},s)$ and if we also assume that the matter is cold---not a useful assumption for neutron star mergers---then we may ignore the entropy and work with a simple barotropic model $e=e(n_\mathrm{b}).$

The arguments we have provided outlines why realistic neutron-star simulations have to be based on a three-parameter equation of state prescription. Reality may be even more complicated, with additional 
particle species and so on,  but the three-parameter model is the minimum we require to describe the hot merger dynamics.

\subsection{Simulations in context}

Having considered the  different assumptions and approximations associated with the matter description we are  well placed to put existing simulations in context. 
In the early days of numerical relativity little attention was paid to the  matter model. The first hurdle to overcome was to simulate matter moving in a dynamical spacetime---a challenging problem in itself. As a result, early simulations were based on barotropic single-parameter models (polytropes being particularly common). As these simulations were (eventually) demonstrated to be robust, the first efforts to account for thermal aspects were made. Much of this work took the barotropic model as starting point and introduced an effective model for the internal energy and the heat. As this approach is still used in simulations (see for example \cite{2020PhRvD.102d4040X,2020PhRvD.101f4052D}), it is useful to explain the logic involved. 

Noting that the barotropic model leads to 
\begin{equation}
p = n_\mathrm{b} \mu_\mathrm{n} - e,
\end{equation}
we may introduce 
\begin{equation}
\mu_\mathrm{n} = m+ \overline\mu,
\end{equation}
where $m$ is the baryon mass.  In terms of the mass density $\rho = m n_\mathrm{b}$, we may then write
\begin{equation}
p =  \rho +  (n_\mathrm{b}\overline\mu- e) =  \rho(1- \epsilon),
\label{peps}
\end{equation}
where $\epsilon$ represents the (specific) internal energy. Finite temperature effect can now be encoded in $\epsilon$. Phenomenologically (see \cite{PhysRevC.92.025801} for a more detailed argument),  comparing to the ideal gas law 
\begin{equation}
p = n k_B T, 
\label{ideal}
\end{equation}
where $k_B$ is Boltzmann's constant (later we use units such that $k_B=1$), we see that $\epsilon = C_v T$, with $C_v$ the specific heat capacity (at fixed volume) while Mayer's relation
\begin{equation}
\frac{k_B}{C_v} = m(\Gamma-1),
\end{equation}
with $\Gamma$ is the adiabatic index, leads to
\begin{equation}
p = \rho \epsilon(\Gamma-1).
\label{gamma}
\end{equation}
This argument motivates the so-called Gamma-law equation of state \cite{2013rehy.book.....R}. It has the advantage of being easy to implement, but  there is no reason to expect it to be particularly realistic---at least not for neutron star mergers. 

The next step up in complexity involves replacing algebraic models like the polytrope with a tabulated equation of state for cold nuclear matter in equilibrium. Tabulated data must be used as the microphysics calculations required to evaluate the equation of state at any given point are prohibitively expense to use live, hence interpolation within a pre-calculated table becomes a necessity. However, one has to pay careful attention because the interpolation has to be thermodynamically consistent in order to avoid modelling errors. The end result is a fairly significant increase in computational cost (as well as memory footprint, which impacts on the potential use of GPUs for large scale simulations). The most simplistic tabulated equation of state tables are barotropic, however many modern simulations (see \cite{Figura_2021} for references to the recent literature) use a cold tabulated equation of state in combination with a Gamma-law prescription to approximate thermal effects. The table then includes pressure $p_\mathrm{cold}\left(\rho\right)$ and specific internal energy $\epsilon_\mathrm{cold}\left(\rho\right)$ from a zero-temperature nuclear physics derived equation of state, and is augmented with a thermal contribution given by
\begin{align}
    p_\mathrm{th}\left(\rho,\epsilon\right) &= \rho \epsilon_\mathrm{th}\left(\rho,\epsilon\right)(\Gamma_\mathrm{th}-1),
\end{align}
where $\epsilon_\mathrm{th}$ is the thermal component of the specific internal energy $\epsilon_\mathrm{th}\left(\rho,\epsilon\right) = \epsilon - \epsilon_\mathrm{cold}\left(\rho\right)$. The total pressure is given by the sum of the cold and thermal components. While this approach is not expected to be realistic (see the discussion in \cite{PhysRevD.102.043006}), it is definitely a step in the right direction.

The key step toward realism involves working with a true finite temperature equation of state. There are a number of such models on the market \cite{CompOSE}, typically developed for supernova simulations. The implementation of these models involves a more complex inversion from evolved to ``primitive'' variables (as we  discuss in \cref{Sec:CodeModifications}) but it does not add conceptual issues, unless we consider problems involving phase transitions \cite{2020EPJA...56...59M,2021arXiv210607885P} or possible mixed matter phases.

When it comes to the matter, the simplest option is to still insist on local equilibrium. The question is if this is a reasonable representation of reality. Are the nuclear reactions fast enough for the equilibrium assumption to be appropriate? Estimates suggest that the answer may be no \cite{2018PhRvL.120d1101A}---at least in parts of the simulated domain---so we need to consider deviations from equilibrium. This raises the complexity level as we need to keep track of the local nuclear reaction rates and implement the associated changes in composition as the simulation proceeds. 
As an alternative, we may consider the slow-reaction limit, in which the composition is effectively frozen. In the case of npe-matter, this involves advecting the lepton fraction along with the fluid flow. This is the assumption we make in our simulation. This may not be an appropriate assumption either, but it is a useful starting point for discussions of out-of-equilibrium issues.  We need to do better, but (as we will argue in the following) there are important lessons to be learned already at this level, and we need to pay attention to these lessons when developing the next generation of simulations.     

\subsection{Going further}

The discussion to this point has assumed a single fluid mixture. That is, the different particle species are assumed to move together with a single four velocity, and in addition the fluctuations in each species must relax to the average on negligible scales \cite{2021arXiv210701083C}. However, there is a range of regions in spacetime and phase space relevant for neutron stars where these assumptions are expected to fail. When dealing with neutrinos the fluid approximation fails. When dealing with superfluidity the single-fluid approximation fails \cite{2021LRR....24....3A}. When considering transport properties, or with phase transitions, quantitative calculations must be done outside the single fluid limit, and the application of these calculations on top of single fluid simulations needs care to ensure the approximations remain reasonable. As an example we may consider the estimate for the thermal conductivity from \cite{2018PhRvL.120d1101A}, which suggests a relaxation timescale of order
\begin{equation}
    t_\mathrm{relax} \sim \left( {\frac{z}{\mathrm{km}}}\right)^2 \left({\frac{T}{10~\mathrm{MeV}}}\right)^2 \ \mathrm{s},
\end{equation}
where $z$ represents the scale of thermal gradients. This rough estimate suggests that gradients on a scale of a few tens of meters would relax on about a millisecond, a timescale similar to that of the anticipated post-merger oscillations \cite{2012PhRvL.108a1101B,2015PhRvL.115i1101B,2016PhRvD..93l4051R}. This estimate is interesting because; on the one hand, it suggests that we could probably get away with ignoring the heat flux (the inclusion of which would require us to go beyond the perfect fluid assumption) on the timescale of a typical merger simulations (especially noting that thermal gradients of order 10 meters may not be resolved in the simulation anyway). On the other hand, the timescale resolvable by a typical simulation is of order a fraction of micro-second (combine a grid resolution of $\Delta x \sim 100$~m with a typical timestep determined by $\Delta t / \Delta x \simeq 0.1$ to see that the resolvable time variation is of order $\Delta t\sim 10^{-7}$~s). That is, the  thermal relaxation may take place on a timescale that should (at least in principle) be resolvable. The issue of small scale thermal features is also closely linked to questions of turbulence and the need to consider a large-eddy approach to the simulations in the first place \cite{2017ApJ...838L...2R, 2021arXiv210701083C}.

To highlight the differences, in a multifluid context the expression for the first law extends from equation~\eqref{eq:FirstLawFull} to
\begin{equation}
    d e = \Theta_a d s^a + \sum_{\text{x}} \mu^{\text{x}}_a d n_{\text{x}}^a,
\end{equation}
where the chemical potentials have extended to conjugate momenta $\mu^{\text{x}}_a$ (with $\Theta_a$ for the entropy, x the species label and $a$ a spacetime index) and the particle number extends to the fluxes $n_{\text{x}}^a$ \cite{2021LRR....24....3A}. In situations where the different fluxes are not aligned, substantial changes in the equilibrium state can be driven by, for example, entrainment. It is important to note that these differences are expected, and may in fact be essential, on some scales. For example, on microscopic scales the difference in charged species flows drives the charge current and hence is required for the neutron star magnetic field. Similarly, on larger scales, the difference between fluxes is central to the hydrodynamical description of superfluids. 

A fully quantitative discussion of transport properties and the interaction of different phases of matter within a neutron star will need to start from a micro-scale theory, such as a multi-species Boltzmann equation based model \cite{2018ASSL..457..455S}. However, the dimensionality of these kinetic-theory based calculations make them computationally prohibitive in many dynamical situations. To estimate to what extent the (single) fluid approximation is valid, the interaction terms between and within each species need to be estimated. Unfortunately, once the fluid approximation is made the information required to confirm its validity is lost, and hence the estimates must be used with care. 

Having outlined how different equation of state models are used in numerical simulations and sketched some of the main points of concern, let us return to the  elephant in the room---the neutrino. It is well known that the neutrinos are central to any discussion of hot neutron star merger dynamics: They remove energy from the system, which may drive matter outflows, and facilitate the reactions that lead to a changing composition which may, in turn, lead to dissipation acting on the fluid dynamics. Yet, current simulations tend to either not include the neutrinos at all or account for them approximately through some (suitably simple) leakage scheme. In fact, many recent discussions \cite{2019EPJA...55..124P,Endrizzi2020} are based on post-processing simulation results. The key point is that the neutrinos do not naturally  lend themselves to a fluid description. Instead, the strategy for dealing with the neutrino problem is formally the same as for photons. The relevant radiation transfer problem must be solved, taking into account frequency dependence and directionality. In current models this is done at different levels of sophistication, starting from some kind of leakage scheme alongside a suitable closure condition for the moment expansion of the radiation stresses (recent examples include \cite{2014MNRAS.441.3177M,2014MNRAS.439..503S, 2018ApJ...860...64F} with, in particular \cite{2020ApJ...900...71A} paying attention to the multi-frequency aspects), increasing realism by considering the angular dependency (as in \cite{2014ApJ...796..106J,2013JCoPh.242..648R}) or taking a  full-blown Monte-Carlo approach (as in \cite{2008JPhCS.125a2007K,2019ApJS..241...30M,2021arXiv210316588F}). In essence,  the problem is complex and extremely expensive computationally.

However, and this is the main point we want to make here, there are issues we need address before we can introduce the neutrino aspects in a faithful fashion. Somewhat simplistically, in order to establish how the neutrinos impact on  the  dynamics we need to know to what extent they are trapped in the simulated ``fluid elements''. This is important as it, in turn, dictates the rate at which weak interactions work to establish local beta equilibrium \cite{Alford2018a,PhysRevD.100.103021}. The issue of deviation from equilibrium comes to the fore and we need to carefully consider what we mean by ``equilibrium" in the first place.

\section{Performing an Out-of-equilibrium Simulation}

As a step towards studying how the equilibrium processes affect the merger dynamics, we have performed a simulation in which the composition of the fluid was advected along with the fluid flow and reactions were ignored. However, before we cover the details of said simulation, let us discuss a few general issues that arise when simulating a continuous fluid on a discrete grid. 

\subsection{Analysis quantities}

Numerical simulations typically provide a limited set of variables which may not be directly useful for extracting physically interesting signatures. Each step in the analysis process that leads to observable or interesting quantities will add numerical error, and may also introduce additional modelling errors in ways that are hard to disentangle.

The key quantities that are controlled in a simulation (based on the standard Valencia formulation, see \cref{Sec:CodeModifications} for a summary of the relevant evolution equations) are the conserved variables---the relativistic equivalents of rest mass density, spatial momenta, and energy. These are represented as cell integral averages in the simulation, and no information about the structure on sub-grid scales is available. Whilst some simulations use large-eddy-style approaches \cite{2017ApJ...838L...2R, 2021arXiv210701083C}, these only give the (assumed) contribution from physics on smaller scales, and not the sub-grid structure itself.

Many quantities of interest can be computed locally using purely algebraic relations, for example thermodynamical quantities from the equation of state. The accuracy of these quantities is limited by the numerical error as well as  modelling errors. The modelling errors  come from, in particular, neglecting sub-grid fluctuations, and hence assuming that the average of the thermodynamical quantity is equivalent to the thermodynamical quantity of the average. This may be reasonable over the bulk of the spacetime, but the modelling error is hard to quantify in the (necessarily under-resolved) turbulent region that forms in the merger process. Notably, this is the hottest region of the simulation.

Additional quantities can be calculated using non-local operations. Examples  include vorticities (using derivatives of given or computed quantities) or optical depths (using integrals of given or computed quantities). In addition to the problems noted above, this will involve additional  error due to the numerical approximation of the non-local operator. Moreover, as non-local operators are typically less accurate at high frequencies (see, e.g.~\cite{leleCompactFiniteDifference1992}), they will again be problematic in the turbulent region of interest. Different quantities of interest may be robust while others are not. For example, the gravitational-wave signal depends on bulk integrated quantities (in the sense of the quadrupole formula) which are likely to be robust as they mainly depend on low-frequency features. In contrast, emission spectra may be less reliable due to the dependence on  high-frequency effects.

Finally, there are quantities that can be considered ``local'' but which depend on non-local constructions. The key example of this is Lagrangian tracers, which have been used to investigate the dynamics of phase transitions and other neutron star properties  \cite{2020EPJA...56...59M,2021arXiv210607885P}. This relies on evaluating the thermodynamic potentials and velocity field at a ``single particle point'' which is then advected through the spacetime. Using a set of point quantities, rather than cell-integral-average quantities, one might  hope to bypass the sub-grid structure issues discussed above. However, all the problems of non-local operators and the modelling of sub-grid structure discussed above will enter through the required interpolation scheme.

This is, in fact, an interesting problem. Lagrangian tracers as an analysis probe of fluid properties have been investigated in depth in the Newtonian literature (example reviews  include \cite{2002AnRFM..34..115Y}). In particular, the discussion in \cite{2002AnRFM..34..115Y} emphasises that three requirements must be met to get accurate results from Lagrangian tracers. First, the simulation must resolve sufficiently small scales in the instantaneous velocity field. This will severely limit the utility of tracers passing through turbulent regions of neutron star simulations. Second, the interpolation scheme used must be of sufficiently high accuracy and differentiability. Finally, a sufficient number of tracers must be used for statistical accuracy (e.g., so that the average Lagrangian velocity is sufficiently close to the average Eulerian velocity). This last point highlights that Lagrangian tracer information should be used statistically.
The analysis in \cite{2002AnRFM..34..115Y} also demonstrates that the number of tracers required scales with the Reynolds number, which may be impractically high for neutron star applications.
These issues highlight why we focus on local grid-cell based analysis in the following.

\subsection{Our Setup}

To take the first step towards quantifying the out of equilibrium behaviour in binary neutron star mergers we focus on a single simulation performed using  the Einstein Toolkit~\cite{EinsteinToolkit}. We used the BSSN formalism of the ADM equations as provided by McLachlan~\cite{McLachlan} for the spacetime evolution and GRHydro~\cite{Baiotti2005} for the fluid evolution. The microphysics is represented by the APR equation of state~\cite{Schneider2019} from the CompOSE database~\cite{CompOSE}. This model can be seen as a ``vanilla'' representation of the relevant physics as it assumes matter composed only of neutrons, protons and electrons and the obtained neutron star radii accord well with current observational constraints (see for example \cite{2021arXiv210506981R}). In order to perform the simulation, it was necessary to make modifications to both GRHydro and EOS\_Omni---the general purpose equation of state interface available in the Toolkit---in order to support the three-parameter tabulated form of the APR equation of state from CompOSE. We describe these modifications in \cref{Sec:CodeModifications}. 

Initial data was obtained using LORENE~\cite{LORENE}, with a fixed temperature of $T=0.02\mathrm{MeV}$ across the entire domain (we comment on this later in \cref{Sec:TempComments}), and under the assumption of cold $\beta$-equilibrium (see \cref{Sec:ColdBetaEquilibrium}). It is important to note that, due to the low-temperature cut-off in the equation of state table we have to assume an initial temperature that is much higher than the expected core temperature of a mature neutron star (below $10^6$~K $\sim 10^{-4}\mathrm{MeV}$ or so). The reason this is an issue will become clear in the discussion of \cref{Fig:APRtauderiv} in \cref{Sec:TempComments}. 
Finally, the initial separation of the two stars was $40~\mathrm{km}$, and each star had an initial baryon mass of $M_\mathrm{b} = 1.4\,M_\odot$. 

While the composition of the fluid is fixed at the initial data stage to be in $\beta$-equilibrium, this constraint was not imposed on the fluid during the evolution, instead the electron fraction $Y_\mathrm{e}$ was advected along with the fluid flow. The simulation does not feature reactions between the constituent species of the fluid, neutrinos are not included, and no magnetic fields are present.

\subsection{Comments on the temperature} \label{Sec:TempComments}

\begin{figure}[bt]
\centering
\includegraphics[width=0.95\textwidth]{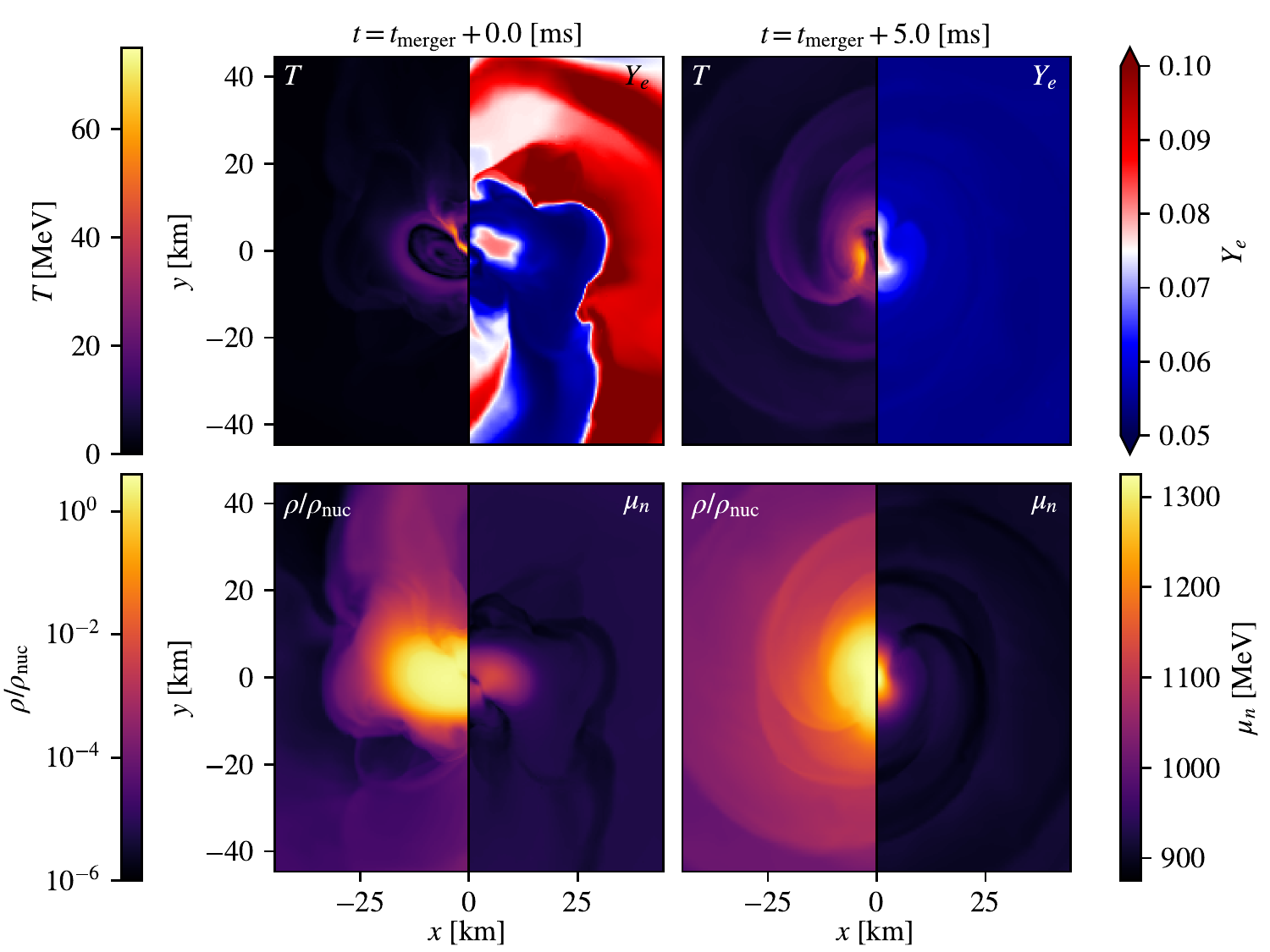}
\caption{\small Temperature $T$, electron fraction $Y_\mathrm{e}$, rest-mass density with respect to nuclear saturation density $\rho / \rho_\mathrm{nuc}$, and neutron chemical potential $\mu_\mathrm{n}$ for a simulation of the merger of two $1.4\,{M}_\odot$ neutron stars using the APR equation of state from \cite{Schneider2019}. The left column coincides with the merger time, while the right column snapshots are $5.0\,\mathrm{ms}$ post-merger.}
\label{Fig:APRrhoTYe}
\end{figure}

At a glance, the results of our simulation agree well with other models considered in the literature, see for example \cite{2019EPJA...55..124P,Endrizzi2020,2021arXiv210607885P}, and we believe they provide a fair representation of current simulation technology. The results we discuss should not, in any sense, be ``particular'' to our chosen model. This is evident if we compare the snapshots in \cref{Fig:APRrhoTYe} to similar results in the literature. 

The results in \cref{Fig:APRrhoTYe} show the
temperature $T$, electron fraction $Y_\mathrm{e}$, rest-mass density with respect to nuclear saturation density $\rho / \rho_\mathrm{nuc}$, and the neutron chemical potential $\mu_\mathrm{n}$ in the equatorial plane, both at the time of merger (left panels) and 5~ms later (right panels). The results are relevant for a range of questions one may want to answer with simulations. For example, while the shock heating associated with the merger heats the matter up to temperatures well above 10~MeV, the high-density core of the merger remnant remains relatively cold---the post-merger configuration is effectively a rotating, fairly cold, high-density ``peanut'' that lasts for some time (we have not tracked the evolution through to the stage where the remnant may eventually collapse for form a black hole as our main interest is in the dynamics immediately following the merger). 

The temperature results impact on the expected post-merger oscillations and issues relating to, for example, bulk viscosity (which we will return to in \cref{sec:bv}). It is notable that the thermal hotspots, which start out localised and co-rotating with the high-density core, spread out as the evolution proceeds. Finally, the results for the chemical potential demonstrate that the matter outflow involves relatively large values of $\mu_\mathrm{n}$, while the values for the electron fraction are moderate. These features influence the r-process (with lower values of $Y_\mathrm{e}$ required to form heavier elements) and the signature of the anticipated kilonova emission. It is also worth noting the spiral nature of the matter outflow, involving colder matter associated with the rarefaction wave following the merger shock. This feature is notable because it is yet another aspect that is affected by the low-temperature cut-off in the equation of state table. It is important to get this part right as it impacts directly on both the r-process and possible jet formation. There are (obviously) many interesting issues here, but we  focus on questions relating to the temperature and the matter composition in the following.

\begin{figure}[tb]
\centering
\includegraphics[width=0.95\textwidth]{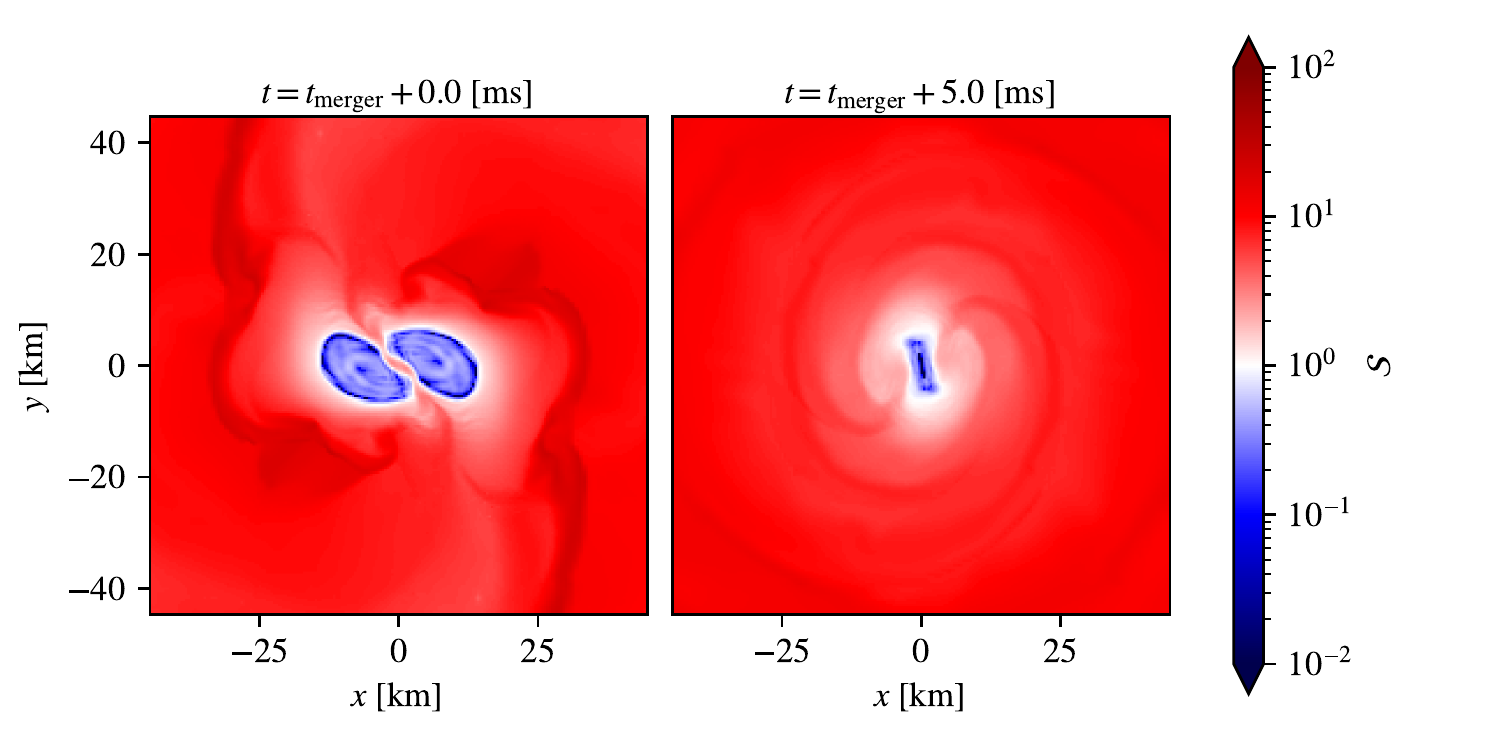}
\caption{\small Total entropy per baryon $\mathcal{S}$ for the simulation of the merger of two $1.4\,{M}_\odot$ neutron stars using the APR equation of state from \cite{Schneider2019}. The left panel coincides with merger time, while the right panel is $5.0\,\mathrm{ms}$ post merger. Note that $\mathcal{S}\leq 1$ indicates degenerate matter while $\mathcal{S}\gg 1$ indicates strongly non-degenerate matter.}
\label{Fig:APREntropy}
\end{figure}

Given this particular focus, it is worth remarking on the robustness of the numerical implementation. As a measure of the reliability of the thermal description we may consider the total entropy per baryon, as shown in \cref{Fig:APREntropy}. This is a useful  measure  (as it should not be negative) which provides a good idea of to what extent the high-density matter is hot or cold (in a nuclear physics sense). The entropy results bring out the expectations---the high density core remains cold---but they also highlight an artefact of the simulations (also apparent in, for example, the results of \cite{2019EPJA...55..124P,Endrizzi2020,2021arXiv210607885P}). The region close to the neutron star surface is artificially hot already from the early stages of the simulation. This is  also apparent from the temperature result in \cref{Fig:APRrhoTYe}. We see that, away from the shock region (where the two stars first touch) the low density matter near each star's surface has reached a temperature well above 10~MeV. This feature is clearly ``unfortunate'' as it means that we cannot discuss the fine print of the thermal physics (e.g. the point at which the neutron star crust melts).

The artificially high surface temperature may, perhaps intuitively, be explained in terms of a numerical shock associated with the transition from the high-density neutron star interior to the low-density (artificial) atmosphere.  Consider any perturbation in the interior that leads to a wave propagating outwards. As it travels ``down'' the density gradient, the speed of sound (and hence the characteristic speed) decreases, meaning that the perturbation steepens and will eventually lead to a shock. There is an assumption that this  shock takes place before hitting the atmosphere (a relevant toy problem demonstrating the effect is discussed in \cite{2011JFM...676..237G}), which seems to match the numerical results, but the subsequent propagation through the atmosphere is certainly artificial. Such a feature may (to some extent) be unavoidable. One may argue that this is not a major problem, as the artificial heating is overwhelmed by the entropy generated in the merger and therefore does not feature prominently in the post-merger dynamics. Pragmatically, this may be true on average and hence sufficient for gravitational wave emission (which is determined by the bulk matter dynamics), but may not be true for determining correct local matter properties (which is needed for e.g.\ electromagnetic and neutrino emission). One would also like to be able to distinguish the (fairly large) modelling error due to this artificial heating from the robust merger physics. 

The surface artefact warrants closer scrutiny since the feature may cause problems for more the detailed modelling of the neutrino aspects and the associated weak interactions.  In particular, the temperature is required to decide if neutrinos are trapped (a point we return to in \cref{sec:neut}). The artificial surface feature may ``confuse'' any automated neutrino treatment and hence it would seem important to understand if this makes a---qualitative or quantitative---difference. 

It is also worth noting that there may  be a link to the initial data prescription. Our initial model assumes a uniform temperature, which would not be a realistic representation of a mature neutron star (more likely represented by a uniform redshifted temperature). The choice is due to the initial data construction within LORENE, which requires a barotropic model. The immediate alternative to our choice would be to assume  a uniform entropy distribution, but this does not represent the expected physics either. Recent work \cite{2019PhRvD.100l3001C} has  considered more realistic entropy profiles, but these have not yet been used in merger simulations. We are investigating this issue in more detail but are not yet in a position to comment on it further. 

\begin{figure}[tb]
\centering
\includegraphics[width=0.95\textwidth]{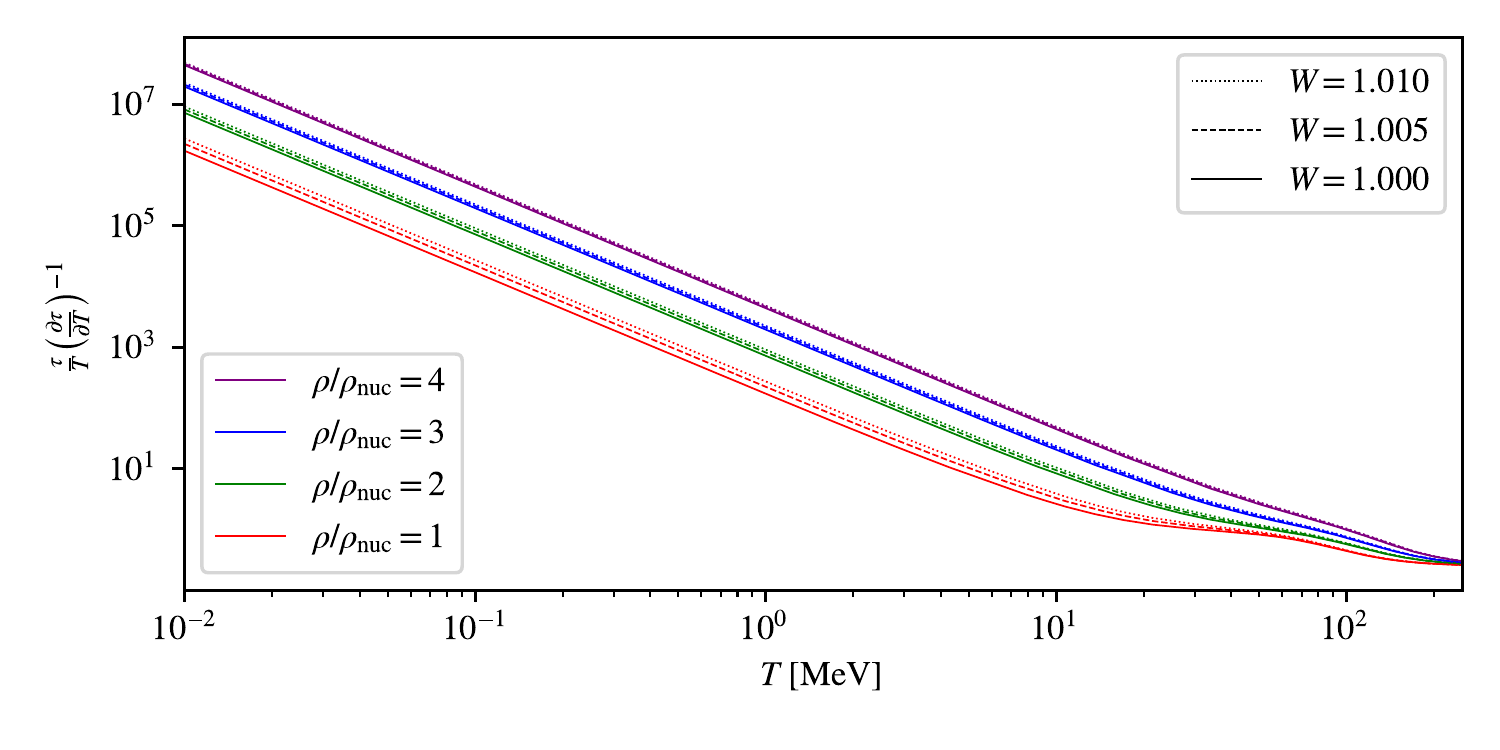}
\caption{\small The condition number ${\cal K}_{\tau \to T} = \frac{\tau}{T} \left(\frac{\partial \tau}{\partial T}\right)^{-1}$ which illustrates how an intrinsic numerical error in the energy $\tau$ is amplified to an error in the temperature. The behaviour is robust to changes in density and velocity (as encoded in the Lorentz factor $W$ explicitly defined in~\cref{Sec:GRHydroMod}). For the APR equation of state from \cite{Schneider2019} we see the result generically diverges as the temperature decreases, indicating serious problems with numerical accuracy at low temperatures.}
\label{Fig:APRtauderiv}
\end{figure}

Finally, it is worth commenting on the temperature of the relatively cold high-density region. As is evident from the snapshots in \cref{Fig:APREntropy}, the temperature in the core of the merger remnant is at the level of 5~MeV. This may be a reasonable reflection of the physics but the result also has to be considered with caveats. As the evolved internal energy $\tau$ only depends weakly on the temperatures (see \cref{Sec:CodeModifications} for a discussion of the evolved variables), the numerical inversion to extract the temperature may be associated with significant uncertainties.
We can approximately quantify this by a \emph{condition number} ${\cal K}_{A \to B}$ which gives, to first order, the relative error in $B$ given a relative error in $A$. For the temperature, we assume an error $\Delta \tau$ in the energy $\tau$, which induces an error in $T$ as
\begin{align}
    \frac{\Delta T}{T} &= \frac{1}{T} \left(\frac{\partial \tau}{\partial T}\right)^{-1}  \Delta \tau \nonumber \\
    &= \left[ \frac{\tau}{T} \left(\frac{\partial \tau}{\partial T}\right)^{-1} \right] \frac{\Delta \tau}{\tau} \nonumber \\
    &= {\cal K}_{\tau \to T} \frac{\Delta \tau}{\tau}.
\end{align}
We then compute ${\cal K}_{\tau \to T}$ for a range of densities, velocities, and temperatures, leading to the results shown  in~\cref{Fig:APRtauderiv}.

We see that the condition number, indicating the growth in the inevitable numerical error, is largely insensitive to the density and velocity. It is, however, extremely sensitive to the temperature and grows rapidly as the temperature decreases. Whilst condition numbers only give a qualitative, rather than quantitative indication, this  suggests that working directly with the temperature will be prone to large numerical errors. This is not a feature of all numerical operations converting between conserved and primitive variables, as the analogous condition number for the densities ${\cal K}_{D \to \rho} = 1$.

If we want to do (significantly) better, we may have to work with a different set of variables (e.g.\ the entropy). This kind of development may be required if we want to be able to explore astrophysically motivated questions associated with the neutron star crust and high-density condensates, which come into play at temperatures below about 1~MeV. However, as it would either involve a different equation of state parameterisation, or dealing with non-conserved fluxes, it is unlikely to be a preferred immediate direction of travel. 

\begin{figure}[bt]
\centering
\includegraphics[width=0.95\textwidth]{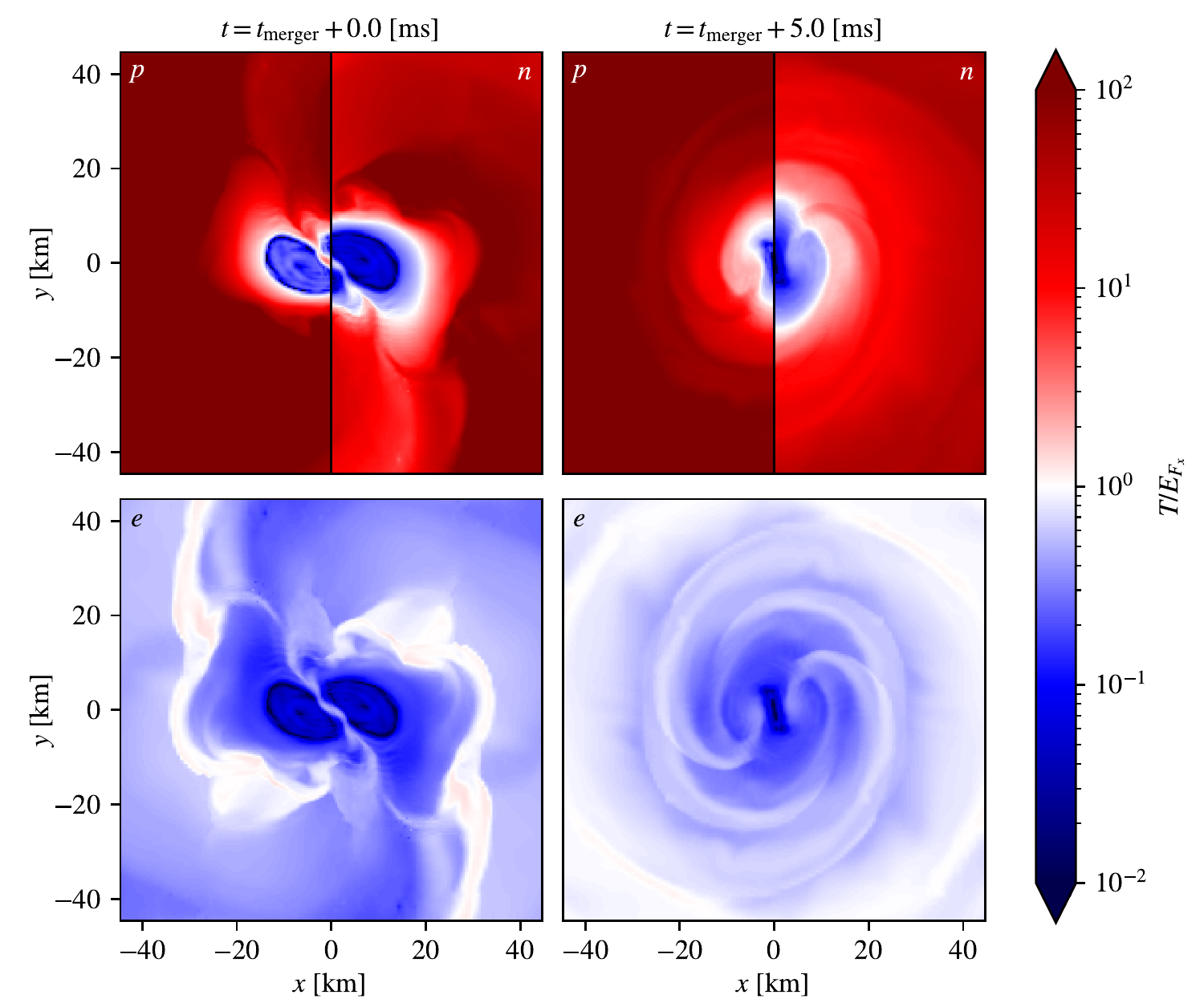}
\caption{\small Degeneracy parameter $T/E_{F_\mathrm{x}}$ for protons $p$, neutrons $n$, and electrons $e$. The baryons are assumed to be non-relativistic while the electrons are assumed to be relativistic. The data relate to the simulation of the merger of two $1.4\,{M}_\odot$ neutron stars using the APR equation of state from \cite{Schneider2019}. The left column coincides with merger time, and right column is $5.0\,\mathrm{ms}$ post merger. Note that $T/E_{F_\mathrm{x}}\ll 1$ indicates strongly degenerate matter, and $T/E_{F_\mathrm{x}}\gg 1$ indicates strongly non-degenerate matter.}
\label{Fig:APRdegen}
\end{figure}

The simulation results also allow us to comment on issues related to the finite temperature equation of state. In particular, we may consider the degeneracy of the different matter components. This is crucial, because the evaluation of the  thermodynamic integrals required to build the equation of state in the first place is relatively ``straightforward'' for both degenerate matter (where we are essentially dealing with a low-temperature expansion) and highly relativistic matter. In the intermediate regime, the calculation is more involved (see for example \cite{2019EPJA...55...10L}). Considering the results from \cref{Fig:APREntropy,Fig:APRdegen} we see that much of the high density core remains strongly degenerate. However,  as the temperature varies significantly across the merger remnant we always have to consider the (more challenging) cross-over regions for the baryons, especially if we are interested in the low-density matter dynamics. It is useful to keep this in mind.

\section{The issue of equilibrium}\label{Sec:Results}

As we declared from the outset, our aim is not to consider the wider aspects of the neutron star merger problem (like the detectable signatures in different channels,  electromagnetic and gravitational). Rather, we want to focus on the  implementation of the physics. In particular,  we want to better understand which aspects of the rich physics (that we would ideally like to  probe) are realistically within reach of a simulation and which ones are not? This may seem a somewhat conservative strategy given that recent simulation work considers a much wider range of issues (matter outflows and nucleosynthesis, electromagnetism and jet launch leading to observed gamma-ray bursts, neutrino transport and so on). However, as will become evident, important questions remain to be resolved already at this more ``basic'' level. In particular, we will focus on the issue of local equilibrium, an issue of immediate relevance for problems involving nuclear reactions (from bulk viscosity to neutrino emission).

\subsection{Cold Chemical Equilibrium}
\label{Sec:ColdBetaEquilibrium}

It is well known that $\beta$-equilibrium in neutron star matter is dictated by a balance of the different electron capture (ec in the following) and neutron decay (nd) reactions. Schematically, these take the form
\begin{align}
    p + e^- + \ldots &\rightarrow n + \ldots, \qquad & \mbox{(ec)} \label{Eqn:ElectronCapture} \\
    n + \ldots &\rightarrow p + e^- + \ldots, \qquad & \mbox{(nd)} \label{Eqn:NeutronDecay}
\end{align}
where one of the ellipses for each reaction contains a neutrino, and they may both contain a spectator nucleon to ensure momentum balance. When the fluid is cold, it is transparent to neutrinos, and thus they may only appear as products for each reaction. In fact, because of their long mean-free path, the neutrinos depart the systems as soon as they are formed. This leaves us with a total of 6 reactions that may take place at $T \rightarrow 0$: The direct Urca processes
\begin{align}
    p + e^- &\rightarrow n + \nu_\mathrm{e}, \label{Eqn:dUEC} \\
    n &\rightarrow p + e^- + \overline{\nu}_\mathrm{e}, \label{Eqn:dUND}
\end{align}
and the modified Urca processes (which require a bystander nucleon)
\begin{align}
    p + p + e^- &\rightarrow n + p + \nu_\mathrm{e}, \label{Eqn:mUpEC} \\
    n + p &\rightarrow p + p + e^- + \overline{\nu}_\mathrm{e}, \label{Eqn:mUpND} \\
    n + p + e^- &\rightarrow n + n + \nu_\mathrm{e}, \label{Eqn:mUnEC} \\
    n + n &\rightarrow n + p + e^- + \overline{\nu}_\mathrm{e}. \label{Eqn:mUnND}
\end{align}
Equivalent reactions involving muons will also be involved at sufficiently high temperatures and densities, but we ignore them here for simplicity. Moreover, as the electrons are relativistic across the simulation domain (we have $\mu_\mathrm{e} \gg T$), positron occupation will be suppressed by a factor of $\exp\left(-\mu_\mathrm{e}/T\right)$, so we can safely ignore reactions involving them, as well \cite{Alford2018a}.

To obtain a condition for equilibrium between the relevant reactions, we need a relation of the form
\begin{align}
    N_{\mathrm{i}_1} P_{\mathrm{i}_1} + N_{\mathrm{i}_2} P_{\mathrm{i}_2} + \ldots &\leftrightarrow N_{\mathrm{j}_1} P_{\mathrm{j}_1} + N_{\mathrm{j}_2} P_{\mathrm{j}_2} + \ldots, \label{Eqn:DetailedBalanceReaction}
\end{align}
where $P_\mathrm{k}$ are the particle species, and $N_\mathrm{k}$ are stoichiometric coefficients representing the balance between products and reactants in a given reaction. The principle of detailed balance means that the reactions will be in equilibrium when the chemical potentials obey the relation
\begin{align}
    N_{\mathrm{i}_1} \mu_{\mathrm{i}_1} + N_{\mathrm{i}_2} \mu_{\mathrm{i}_2} + \ldots &= N_{\mathrm{j}_1} \mu_{\mathrm{j}_1} + N_{\mathrm{j}_2} \mu_{\mathrm{j}_2} + \ldots, \label{Eqn:DetailedBalanceCP}
\end{align}
where $\mu_\mathrm{k}$ is the chemical potential of species $P_\mathrm{k}$. However, none of the 6 allowed reactions above are the inverse of any other---as they all must feature the neutrino as a product---so we cannot write the reactions in the form suggested in \cref{Eqn:DetailedBalanceReaction}. Instead, we assume that the neutrinos are kinematically negligible (their energy will be on the order of a few times the temperature~\cite{Haensel1992}) allowing us to pair the electron capture and neutron decay reactions together (effectively assuming that they proceed at the same rate, $\Gamma_\mathrm{nd} = \Gamma_\mathrm{ec}$) to give
\begin{align}
    n &\leftrightarrow p + e^-, \\
    n + p &\leftrightarrow p + p + e^-, \\
    n + n &\leftrightarrow n + p + e^-.
\end{align}
These all have the equilibrium condition
\begin{align}
    \mu_\mathrm{n} = \mu_\mathrm{p} + \mu_\mathrm{e}\ , \label{Eqn:ColdBetaEquilibrium}
\end{align}
which dictates the matter composition at $T=0$ (and holds up to temperatures of around a few hundred keV \cite{Alford2018a}).

\begin{figure}[bt]
\centering
\includegraphics[width=0.95\textwidth]{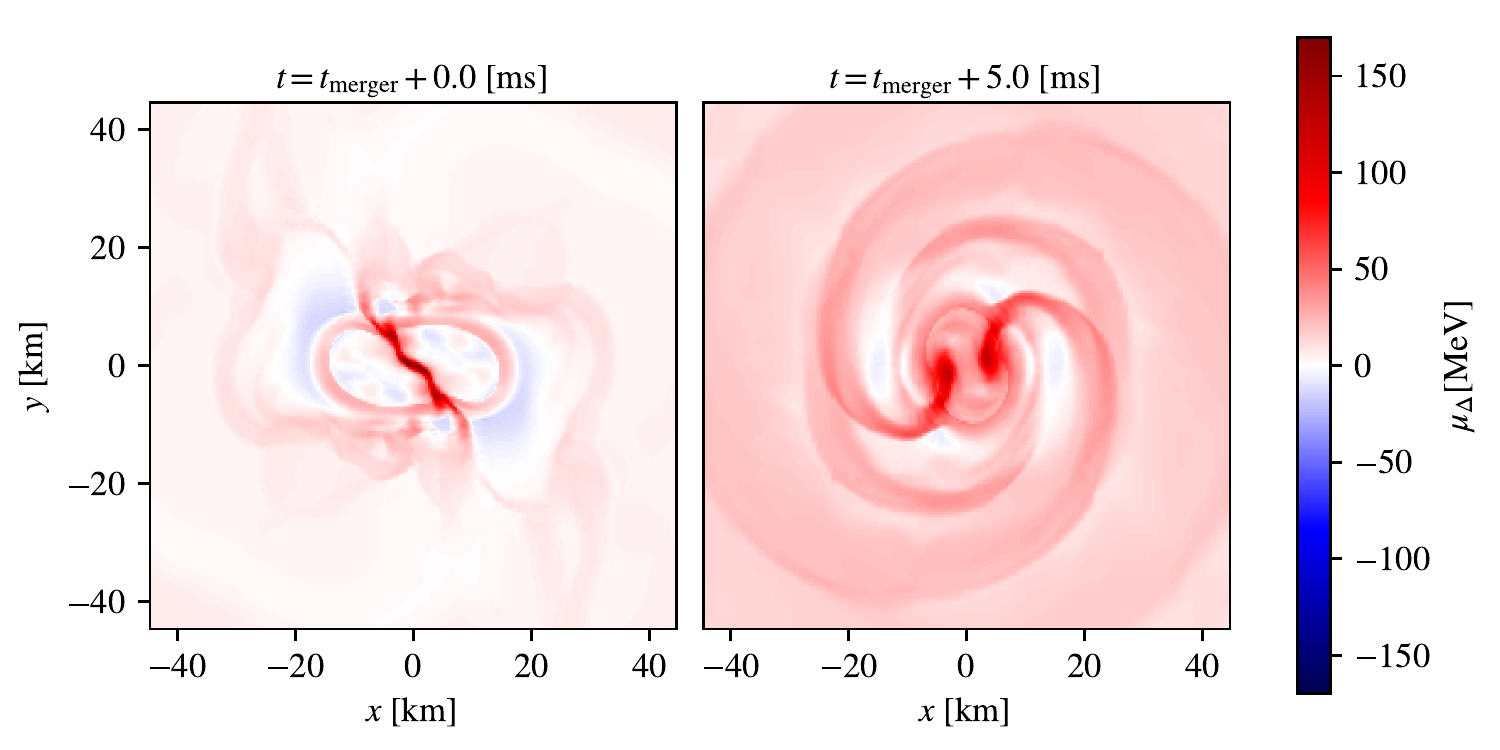}
\caption{\small Deviation from cold $\beta$-equlibrium $\mu_\Delta = \mu_\mathrm{n} - \left(\mu_\mathrm{p} + \mu_\mathrm{e}\right)$ for a simulation of the merger of two $1.4\,{M}_\odot$ neutron stars using the APR equation of state from \cite{Schneider2019}. The left panel coincides with merger time, and right panel is $5.0\,\mathrm{ms}$ post merger.}
\label{Fig:APRMuDelta}
\end{figure}

Given that our simulation does not enforce the equilibrium (the lepton fraction is advected with the fluid), we can use the results to quantify the local deviation from the cold beta-equilibrium. 
To measure this deviation  we introduce 
\begin{align}
    \mu_\Delta &= \mu_\mathrm{n} - \mu_\mathrm{p} - \mu_\mathrm{e}\ . \label{Eqn:EquilibriumDeviationCold}
\end{align}
The results, which are provided (in the equatorial plane) in \cref{Fig:APRMuDelta}, show that there are regions---notably associated with the merger shock and the post-merger hotspots---where $\mu_\Delta \sim 100$~MeV. That is, the matter deviates significantly from the cold equilibrium. Having said that, it is  important to note that the deviation is not uniform throughout the simulation domain.

\subsection{A step towards bulk viscosity}
\label{sec:bv}

\begin{figure}[bt]
\centering
\includegraphics[width=0.95\textwidth]{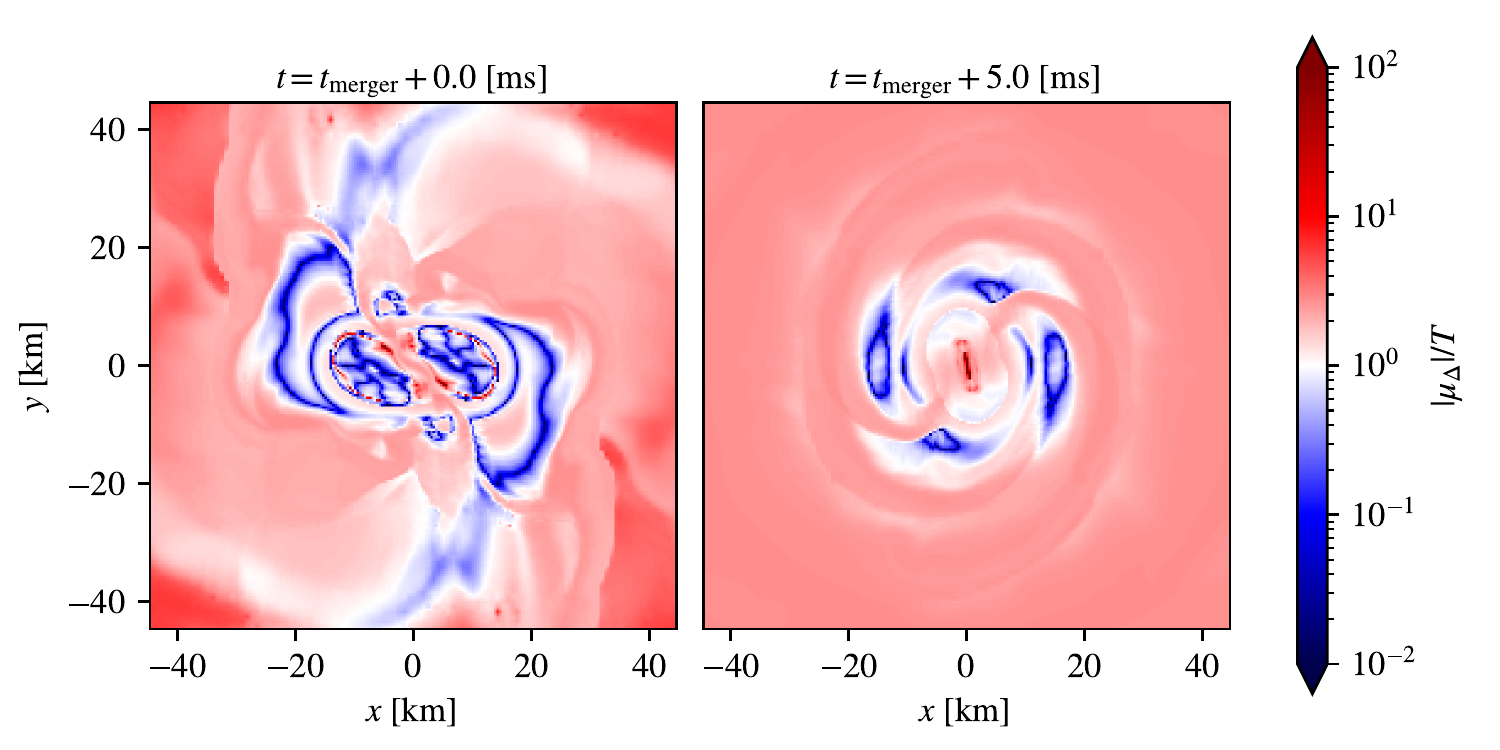}
\caption{\small Bulk viscosity regime parameter $\left| \mu_\Delta \right| / T $ for the simulation of the merger of two $1.4\,{M}_\odot$ neutron stars using the APR equation of state from \cite{Schneider2019}. The left panel coincides with the merger time, and the right panel is $5.0\,\mathrm{ms}$ post merger.  We see that there are regions in the simulation that remain sub-thermal (where a low-temperature expansion and a perturbative analysis of bulk viscosity would suffice), but there are also large regions where the matter is in the suprathermal regime.}
\label{Fig:APRBVRegime}
\end{figure}

The results for $\mu_\Delta$  from \cref{Fig:APRMuDelta} represent a first step towards a discussion of the reactions that serve to reinstate equilibrium and issues relating to, for example, bulk viscosity \cite{Alford2018a,particles3020034}, as they
 allow us to quantify the rate at which the system equilibrates and assess the relevance of bulk viscosity on the fluid dynamics. This argument is notably different from the recent discussion in \cite{2021arXiv210705094M}. Our focus here is on the parameters required to determine the relevant equilibration rates, while \cite{2021arXiv210705094M} mainly considers the nature of the fluid flow during merger. The latter argument is  complicated as 
bulk viscosity is typically thought of as a resonant phenomenon \cite{2018ASSL..457..455S}, which is particularly important when the dynamical motion of the fluid and the rate of relaxation towards equilibrium  occur on comparable timescales. The usual logic is perturbative and the calculation is typically done in terms of a periodic solution for the dimensionless parameter
\begin{align}
    \mathcal{A} = \frac{\mu_\Delta}{T}.
\end{align}
The calculation of the viscosity from $\mathcal{A}$ (see Alford et al. \cite{Alford_2010} for a detailed discussion) can be done in three regimes: a general solution and two limiting cases. First of all, the sub-thermal limit, where $\mathcal{A} \ll 1$, allows an analytic solution, essentially an expansion in small deviations from equilibrium (where it may be sufficient to include only the linear term). Meanwhile, the general solution and the supra-thermal limit, where $\mathcal{A} \gg 1$, are more complicated as they both require numerical integration of the periodic solution. Given this understanding, it is interesting to quantify the regions in the merger simulation where each assumption applies. As an indication of this, we plot in \cref{Fig:APRBVRegime}  the value of $\left| \mu_\Delta \right| / T $ (using \cref{Eqn:EquilibriumDeviationCold}) reached by the matter in our simulation. As far as we are aware, this is the first time this kind of information has been extracted from numerical merger data. The results paint a complex picture. While there are regions in the simulation that remain sub-thermal (where a low-temperature expansion and a perturbative analysis of bulk viscosity would suffice) there are also large regions where the matter is in the suprathermal regime. The upshot is that the analysis of the impact of bulk viscosity in a live simulation would require a careful on-the-fly identification of these regions. 

\begin{figure}[bt]
\centering
\includegraphics[width=0.95\textwidth]{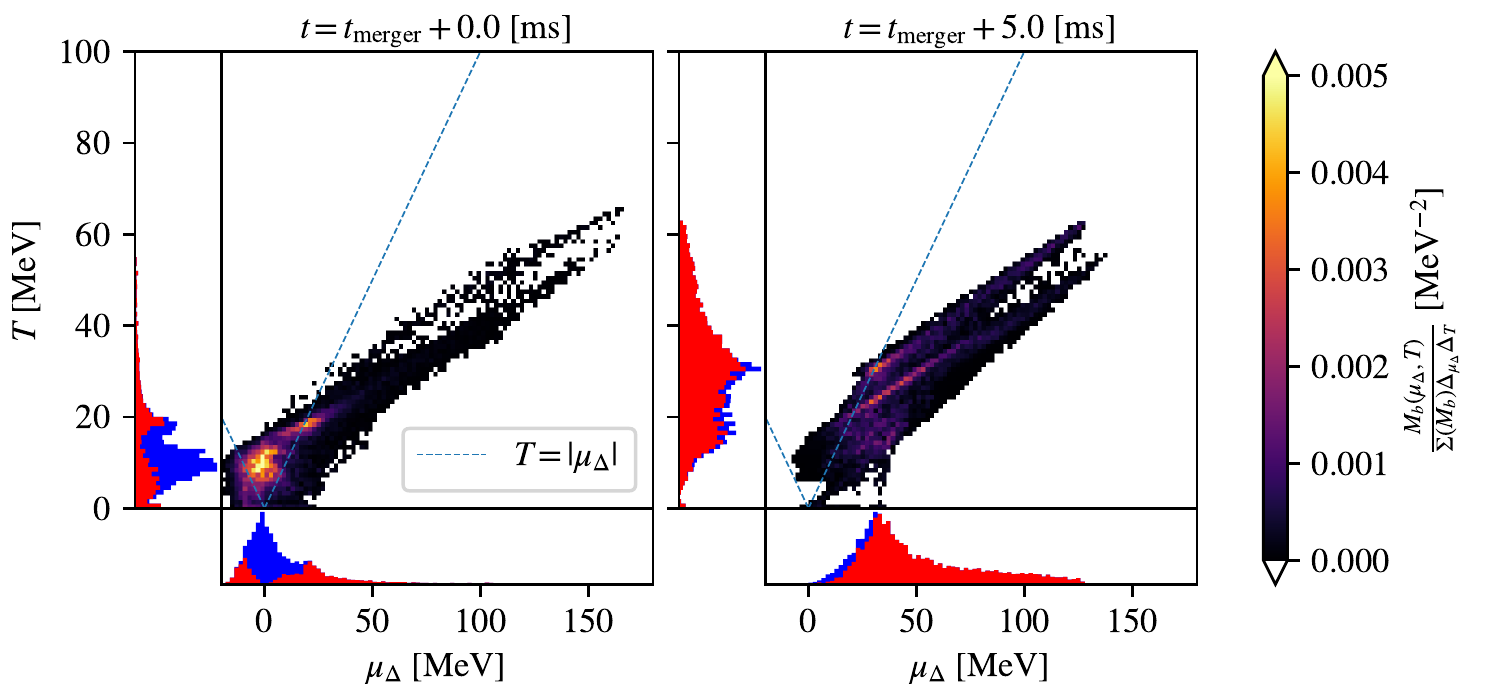}
\caption{\small Distribution of baryon mass $M_\mathrm{b}$ in the $\mu_\Delta$-$T$ plane for a simulation of the merger of two $1.4\,{M}_\odot$ neutron stars using the APR equation of state from \cite{Schneider2019}, where $\mu_\Delta=\mu_\mathrm{n}-\mu_\mathrm{p}-\mu_\mathrm{e}$. The left panel coincides with the merger time, and the right panel is $5.0\,\mathrm{ms}$ post merger. The dashed line denotes $T=\left|\mu_\Delta\right|$. Matter above this line therefore has $\mathcal{A}<1$ and is in the sub-thermal regime, whereas matter below the line has $\mathcal{A}>1$ and is in the supra-thermal regime. This distinction is replicated in the histograms to the side of and below each plot, with blue indicating sub-thermal, and red indicating supra-thermal.}
\label{Fig:APRMuDelta_subsupra}
\end{figure}

We can also look at this in the $\mu_\Delta$-$T$ plane, as we show in~\cref{Fig:APRMuDelta_subsupra}. We see that the majority of the matter reaches the point of merger around $\mu_\Delta=0$, but in the remnant we see that the modal $\mu_\Delta$ shifts to around $\mu_\Delta=40\,\mathrm{MeV}$. There is a strong correlation between $T$ and $\mu_\Delta$, which is perhaps unsurprising as much of the temperature increase in the simulation is driven by compression heating, and changes in density are also able to drive changes in $\mu_\Delta$. In the left panel we see that before merger most of the matter is in the sub-thermal regime, whereas in the right panel most of the matter has crossed the $T=\left|\mu_\Delta\right|$ line to enter the supra-thermal region. While most of this matter is at high temperature, meaning that if reactions were included it would have been quickly driven to a lower $\mu_\Delta$, we see that there is some matter (corresponding to the inner core of the remnant) with $T\lesssim 5\,\mathrm{MeV}$ and $\mu_\Delta\sim 40\,\mathrm{MeV}$, meaning that it may well be necessary to consider supra-thermal bulk viscosity. 

In principle, we could proceed and use the results to estimate the bulk viscosity. There are, however, a couple of issues that prevent us from taking this step. First, as stated in previous section, the calculation of $\mu_\Delta$  assumes that the fluid is at $T \rightarrow 0$ (we are measuring the deviation from the cold equilibrium). As $T \neq 0$ is required for the calculation of $\mathcal{A}$, we need to consider how the problem changes for finite temperatures. The second issue we need to acknowledge relates to the fact that the standard bulk viscosity prescription involves an estimate of the local dynamical timescale. The estimates from \cite{particles3020034} suggest that the bulk viscosity resonance may be relevant for neutrino-transparent matter, but the outcome depends on the local conditions and  the required information is not easily extracted from a live simulation, at least not in a way that does not significantly ramp up the computational cost. A more practical approach would likely involve implementing the relevant reaction network directly, but this is neither cheap nor easy. 

Before we move on, it is worth commenting on the timescales of the reactive problem (as estimated in for example \cite{particles3020034}). 
To outline the argument, following~\cite{2002AnRFM..34..115Y}, consider a general fluid quantity $X$ varying in time and space. Working on the fluid element scale, we can construct Eulerian $\tau_{E, X}$ and Lagrangian $\tau_{L, X}$ timescales for this quantity from
\begin{align}
    \tau_{E, X} &= \left( \frac{\text{Var}(X)}{\text{Var}(\frac{\partial X}{\partial t})} \right)^{1/2}, \\
    \tau_{L, X} &= \left( \frac{\text{Var}(X)}{\text{Var}(\frac{D X}{D t})} \right)^{1/2},
\end{align}
where $D$ represents the appropriate convective derivative. 

It is typical in simulations to note that reactions typically take place on the fast timescale of the weak interaction (of order $10^{-8}-10^{-10}$~s), and hence may be neglected on the timescales that are resolved in the simulation ($\sim 10^{-7}$s). In effect, we assume that the quantity $X$ depends \emph{separately} on the slow timescale $t$ and the fast timescale $\tau_{\text{fast}} = t/(t_{\text{reaction}}/\Delta t)$ and integral-average over the fast timescale. The quantities used in the simulation are then considered to be
\begin{equation}
    \hat{X}(t) = \lim_{T \to \infty} \frac{1}{T} \int_0^T \text{d} \tau_{\text{fast}} \, X(t, \tau_{\text{fast}}).
\end{equation}
When the fluctuations induced by a given reaction decay exponentially on timescales $\sim \tau_{\text{fast}}$ (which follows when their statistics approximate a linear Markov process) then it is reasonable to assume that the reaction takes place ``instantaneously'' and so equilibrium should be imposed on the simulated quantities. When the fluctuations instead are oscillatory there can be ${\cal O}(1)$ differences between $X$ and $\hat{X}$.  This would be the case, for example, for bulk viscosity and it would no longer be appropriate to ignore the fast timescale dynamics.

Moreover, we need to keep in mind that the reactions apply to fluid elements reacting on Lagrangian timescales, while a typical simulation is working in an Eulerian frame on Eulerian timescales. Interestingly, it is noted in \cite{2002AnRFM..34..115Y}  that, in fully resolved Newtonian simulations, these two timescales need not match. In fact, they may be rather different. In turbulent regions with high Reynolds number, the hypothesis is that $\tau_{E, X} / \tau_{L, X} \sim \text{Re}^{1/2}$. This also highlights that the assumptions leading to an imposed, instantaneous equilibrium may not hold in the most interesting regions of a neutron star merger. We clearly need to approach these issues with some care.

\subsection{Equilibrium at Finite Temperature}\label{SecWarmBetaEquilibrium}

Let us return to the issue of equilibrium. 
When working with cold matter, the momenta of the particles taking part in the reactions are assumed to be at their respective Fermi surfaces (this is known as the Fermi surface approximation). At low densities the Fermi momenta $p_F$ of the three involved particles (neglecting the neutrinos) satisfy the relation
\begin{equation}
    p_{F_\mathrm{n}} > p_{F_\mathrm{p}} + p_{F_\mathrm{e}}\ .
\end{equation}
This blocks the direct Urca processes as there is no way to balance the momenta, requiring the introduction of the spectator nucleon in the modified Urca processes. However, as the density increases, $p_{F_\mathrm{p}} + p_{F_\mathrm{e}}$ increases faster than $p_{F_\mathrm{n}}$, and some equations of state allow the direct processes to proceed above a threshold density at which $p_{F_\mathrm{n}} = p_{F_\mathrm{p}} + p_{F_\mathrm{e}}$ \cite{Alford2018a}. For cold matter, this condition typically requires a proton fraction of order 10\% (in the case of the APR model we consider here, the direct Urca threshold is reached at a  around $5\rho_\mathrm{nuc}$, beyond the densities reached in our simulation). 

By relaxing the Fermi surface approximation at finite temperature for the direct processes, one finds that instead of being forbidden, they are Boltzmann suppressed by a factor depending on the single particle free energy $\gamma_\mathrm{i}$, defined by
\begin{equation}
    \gamma_\mathrm{i} (p) = E_\mathrm{i} (p) - \mu_\mathrm{i} = E_\mathrm{i}(p) - E_{F,\mathrm{i}},
\end{equation}
where $p$ is the particle momentum. $\gamma_\mathrm{i} (p)$ therefore represents the energy difference between a particle on the Fermi surface and a particle with momentum $p$. This will result in Boltzmann suppression of the rates of electron capture and neutron decay by a factor of $\exp\left(-\left|\gamma_\mathrm{i}\right|/T\right)$ where each reaction will be dominated by a particular $\gamma_\mathrm{i}$. For neutron decay, the dominant factor is finding a hole in the electron Fermi sea below $p_{F_\mathrm{e}}$ by an amount in the region of $\gamma_\mathrm{e} = 20-25\,\mathrm{MeV}$, whereas for electron capture the energy mismatch is lessened by the anti-aligning of the neutrino produced with the resultant neutron, reducing the increase of momentum needed on the initial proton to around $\gamma_\mathrm{p} = 10-15\,\mathrm{MeV}$ \cite{Alford2018a}. 
The upshot of this is that the reactions are no longer in balance when \cref{Eqn:ColdBetaEquilibrium} holds. In fact, as the direct electron capture rate is suppressed to a lesser extent than the neutron decay rate, there will be a net production of neutrons when the cold equilibrium condition is satisfied. To account for this, one may introduce a chemical potential offset $\mu_\delta$, such that the fluid reaches its actual equilibrium when \cite{Alford2018a}
\begin{align}
    \mu_\mathrm{n} = \mu_\mathrm{p} + \mu_\mathrm{e} + \mu_\delta. \label{Eqn:WarmBetaEquilibrium}
\end{align}
The offset is determined by balancing the rates of the Urca processes, again setting $\Gamma_\mathrm{ec}=\Gamma_\mathrm{nd}$ but now for a finite temperature. This then allows us to work out the matter composition assuming that the temperature is held fixed.

Alford et al. \cite{Alford2018a} (see also the recent discussion in \cite{alford2021beta}) demonstrate that, for temperatures  of order $10\,\mathrm{MeV}$ the required offset can be as large as $\mu_\delta=20-25\,\mathrm{MeV}$. We get an immediate idea of the impact of this, by comparing to the results for $\mu_\Delta$ from \cref{Fig:APRMuDelta}.  Clearly, there are large regions in the simulation where the suggested effect would be significant, bringing the matter closer to a de facto equilibrium than one might have expected. The effect becomes less important after the merger as the thermal hotspots drive outflows and   significant regions are driven much further ($\gtrsim 100\,\mathrm{MeV}$) from the cold beta-equilibrium. 

\begin{figure}[bt]
\centering
\includegraphics[width=0.95\textwidth]{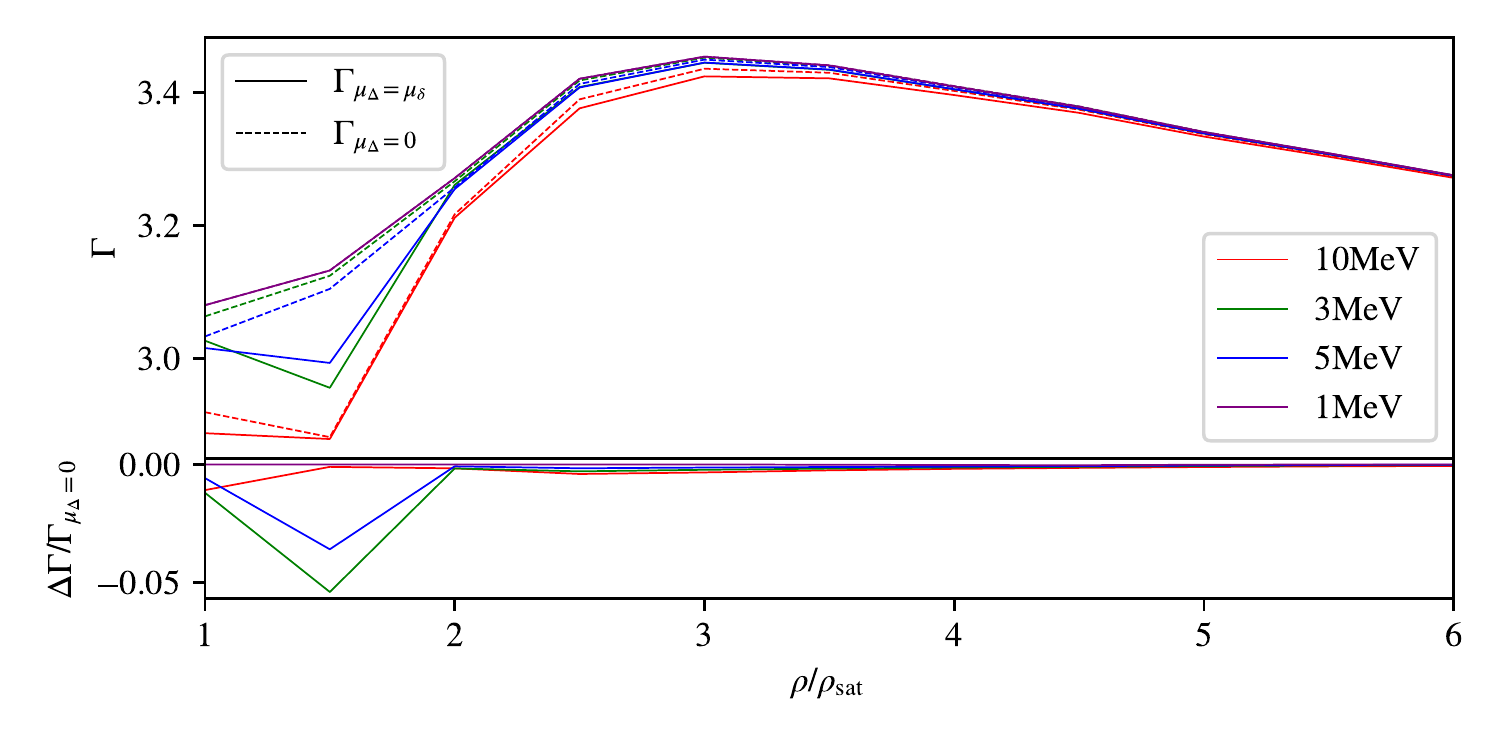}
\caption{\small Illustrating the impact of the warm equilibrium on the stiffness of the equation of state. Dashed lines assume the ``cold'' $\beta$-equilibrium, while solid lines use the ``warm'' equilibrium prescription with $\mu_\delta$ take from Figure~4 in \cite{Alford2018a}. The results show that the effect tends to soften the equation of state by up to 5\% at densities up to $2\rho_\mathrm{nuc}$, but also that the impact is much less pronounced at higher temperatures.}
\label{Fig:APRSoftening}
\end{figure}

Using the results for  $\mu_\delta$ from Alford et al.~\cite{Alford2018a}, we may estimate the degree to which the shift of equilibrium  causes the equation of state to soften in equilibriated matter. This provides a useful insight into the level at which the effect may impact on, for example,  the gravitational-wave signal.  In \cref{Fig:APRSoftening} we plot the adiabatic index $\Gamma$ for the APR equation of state and matter in ``cold'' equilibrium ($\mu_\Delta=0$) as well as matter where the suggested temperature effects are taken into account (setting $\mu_\Delta = \mu_\delta$ and using the results from Figure~4 in \cite{Alford2018a}). We  see that there is a general trend towards lower values of $\Gamma$ as the temperature increases, and the softening effect due to composition changes is most pronounced at intermediate temperatures, peaking at around the $5\%$ level for matter in the $1-2\rho_\mathrm{nuc}$ density range. While we cannot quantify the extent to which this degree of softening impacts on observables, like the frequency of the post-merger oscillations, it is clear that this is a question worth returning to in future work (which would require an equation of state accounting for the temperature dependent $\mu_\delta$). Finally, it is worth noting that the softening associated with the warm equilibrium  is much less pronounced at a temperature of $10\,\mathrm{MeV}$  (and above). This is likely due to the general softening associated with the thermal pressure, an effect that is evident in the top panel of \cref{Fig:APRSoftening},  dominating at higher temperatures.

\subsection{Equilibrium in Matter with Trapped Neutrinos}
\label{sec:neut}

The equilibrium relations given in \cref{Eqn:ColdBetaEquilibrium,Eqn:WarmBetaEquilibrium} both assume that the fluid is transparent to neutrinos, which can then only appear as products in the allowed reactions. However, at sufficiently high temperatures the neutrino mean free path with respect to absorption, $\ell_\mathrm{ab}$, is expected to become short enough that they begin to react with the fluid matter at a meaningful rate, participating in the inverses of the 6 aforementioned Urca reactions. Now that the processes are allowed to proceed in both directions, it follows from \cref{Eqn:DetailedBalanceCP} that all 6 pairs of reactions have the equilibrium condition
\begin{align}
    \mu_\mathrm{n} + \mu_{\nu_\mathrm{e}} &= \mu_\mathrm{p} + \mu_\mathrm{e} \label{Eqn:HotBetaEquilibrium}
\end{align}
where $\mu_{\nu_\mathrm{e}}=-\mu_{\overline{\nu}_\mathrm{e}}$ is the electron-neutrino chemical potential.

Under these conditions, the relaxation timescale is expected to be (see figure 7 in \cite{PhysRevD.100.103021}) of order $10^{-8}-10^{-10}$~s (faster than any current numerical simulation is likely to be able to resolve). This, in turn, leads to a weakening of the bulk viscosity by several orders of magnitude. As this has an obvious impact on the matter dynamics---the new equilibrium that the matter will evolve towards is rather different from the cold and warm cases---we clearly need to consider the neutrinos. Unfortunately, this is problematic. As the neutrinos were not included in our simulation (or indeed the equation of state in the first place!), we cannot directly measure the neutrino chemical potential. We have to resort to approximations. For example, we get some idea of the likely magnitude of the neutrino chemical potential by assuming $\mu_\nu \sim \mu_\Delta$ and then considering the results shown in \cref{Fig:APRMuDelta}. Still, if we want to do better then we need to account for the neutrinos in the matter description. Similarly, we cannot reliably quantify the rate at which the neutrinos are involved in reactions to determine the regions of the simulation for which \cref{Eqn:HotBetaEquilibrium} is the appropriate statement of $\beta$-equilibrium. However, we can estimate these regions by post-processing (as in \cite{Endrizzi2020}) the absorption opacity $\kappa_\mathrm{ab}$ (which is related to the mean free path $\ell$ through $\ell=1/\kappa$). 

Given a path $\lambda$ of length $L$, the probability of a neutrino being transmitted along that path is
\begin{align}
    P_\mathrm{T} \left( \lambda \right) &= \exp\left( - \kappa_\mathrm{ab} L \right).
\end{align}
Hence, we can estimate the regions where neutrino absorption will be significant by examining the magnitude of $\kappa_\mathrm{ab} L$. In regions where $\kappa_\mathrm{ab} L \gg 1$, the probability of transmission is very low, and thus neutrinos are available to partake in reactions, whereas in the opposite limit they are likely to escape freely. The question then is, what should we take to be the relevant length scales? Pragmatically, in merger simulations there are two pertinent length scales: the grid spacing $\Delta x$, which for the simulation under discussion is $\Delta x \approx 400\mathrm{m}$, and the size of the region containing hot matter $r_\mathrm{hot}$, which here is $\sim 30-50\mathrm{km} \approx 100 \Delta x$, see \cref{Fig:APRrhoTYe}. 

In regions where $\kappa_\mathrm{ab} \Delta x \gg 1$ the neutrinos are unlikely to escape the computational cell in which they were emitted, so for our purposes they are definitely trapped and available for reactions. Similarly, in regions where $\kappa_\mathrm{ab} r_\mathrm{hot} \ll 1 \implies \kappa_\mathrm{ab} \Delta x \ll 1/100$ the neutrinos are likely to escape the simulation without reacting, and so these regions should evolve towards the cold/warm equilibrium conditions on some relatively fast timescale. In the intermediate regime, the neutrinos emitted in one place are not likely to be absorbed locally (i.e. within the same computational cell), but they are also unlikely to escape the simulation completely, hence it is difficult to make a definite statement one way or the other---without directly simulating the neutrinos. These arguments show why the neutrino treatment is problematic. The logic that the matter is either neutrino transparent or not is too simplistic to describe the conditions throughout much of the simulation domain. Nevertheless, as one has to start somewhere, let us consider where the estimates we have suggested take us.

The opacity of the fluid to neutrinos depends on the equation of state parameters $\left(\rho,T,Y_\mathrm{e}\right)$, the neutrino energy $E_\nu$, and the neutrino species $x$. We use NuLib \cite{NuLib} to calculate $\kappa_\mathrm{ab} \left(\rho, T, Y_\mathrm{e}, E_\nu, x \right)$ for the two most relevant species, the electron neutrino/anti-neutrino. In order to obtain an estimate for the appropriate neutrino energies, we assume that there are sufficient neutrinos available to be in chemical equilibrium, and that those neutrinos will be in thermal equilibrium with the fluid. Endrizzi et al.~\cite{Endrizzi2020} have shown that this assumption is good enough to determine equilibrium surfaces, which depend strongly on the absorption, and thus this approximation should work well for the qualitative analysis here. The neutrinos then  follow an isotropic Fermi-Dirac distribution
\begin{align}
    f_\nu \left(E_\nu\right) &= \left[ \exp\left( \frac{E_\nu - \mu_\nu}{T} \right) + 1 \right]^{-1},
\end{align}
where $\mu_{\nu_\mathrm{e}} = -\mu_{\overline{\nu}_\mathrm{e}} = \mu_\mathrm{p} + \mu_\mathrm{e} - \mu_\mathrm{n}$. We then calculate an energy averaged opacity $\tilde{\kappa}_\mathrm{ab}$ for each species through (as we focus on particle number transport~\cite{Endrizzi2020})
\begin{align}
    \tilde{\kappa}_\mathrm{ab} \left(\rho, T, Y_\mathrm{e}, x \right) &= \frac{\displaystyle \int^{\infty}_{0} f_\nu\left(E_\nu\right) \kappa_\mathrm{ab} \left(\rho, T, Y_\mathrm{e}, E_\nu, x \right) E^2_\nu \mathrm{d} E_\nu}{\displaystyle \int^{\infty}_{0} f_\nu\left(E_\nu\right) E^2_\nu \mathrm{d} E_\nu}.
\end{align}

\begin{figure}[bt]
\centering
\includegraphics[width=0.95\textwidth]{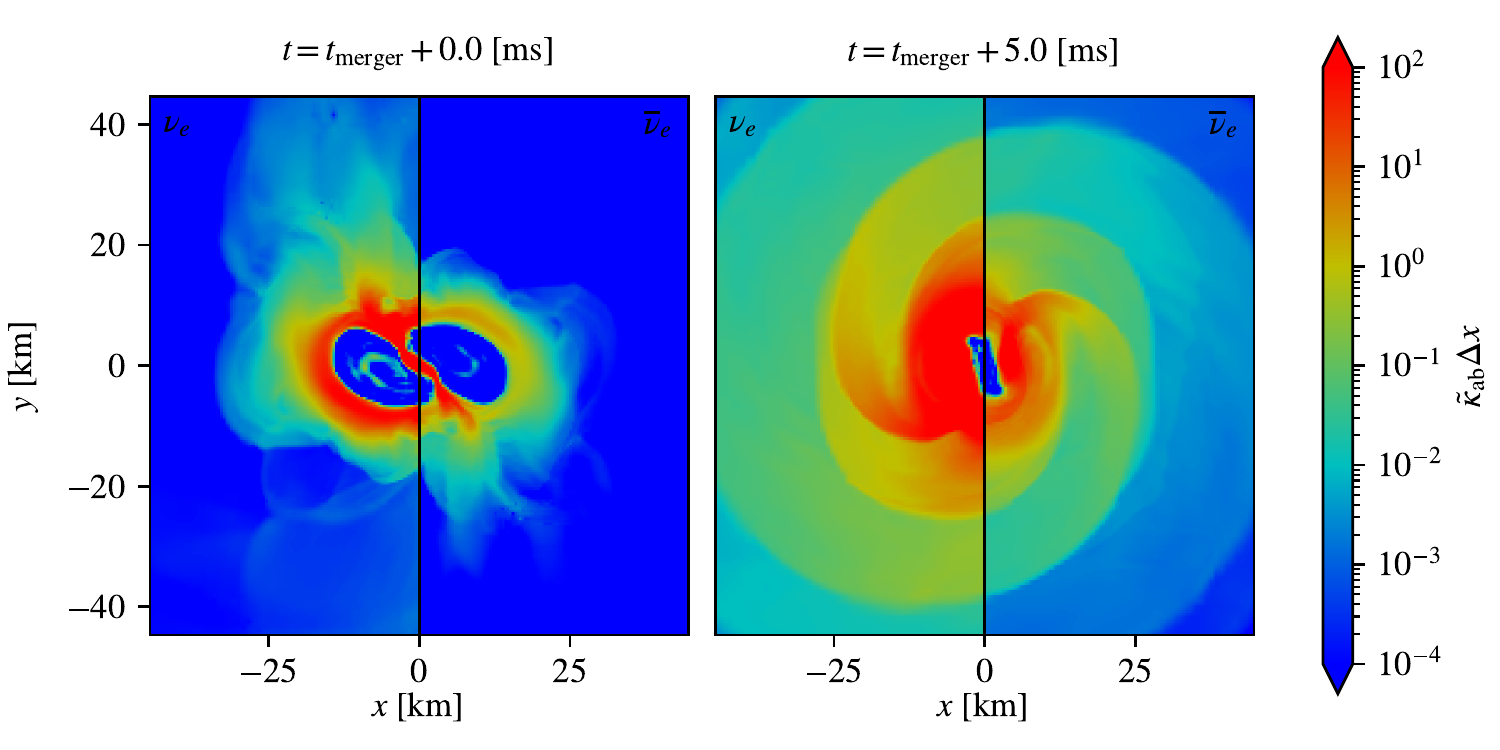}
\caption{\small Energy averaged electron neutrino and anti-neutrino absorption opacities $\tilde{\kappa}_\mathrm{ab}$ multiplied by grid cell spacing $\Delta x = 400\,\mathrm{m}$ for a simulation of the merger of two $1.4\,{M}_\odot$ neutron stars using the APR equation of state from \cite{Schneider2019}. The left panel coincides with the merger time, and the right panel is $5.0\,\mathrm{ms}$ post merger. The neutrino opacities were calculated using NuLib~\cite{NuLib}.}
\label{Fig:APRkab}
\end{figure}

The results in \cref{Fig:APRkab} show that $\tilde{\kappa}_\mathrm{ab} \Delta x$ is small in the dense core regions up to merger, and although some of the core matter in the remnant has been heated enough that absorption on the scale of grid cells has become significant, there remains a transparent central region. The hotspots mentioned in \cref{Sec:TempComments} show large potential for absorption of both species of neutrino, as expected. The outflowing matter visible in the $+5\,\mathrm{ms}$ panel is mostly in the intermediate regime, however it does highlight the difference between the values of $\tilde{\kappa}_\mathrm{ab}$ for the two different neutrino species, with the electron neutrinos being subjected to a stronger degree of absorption than their antiparticle. Hence we are likely to see regions where the inverses of the Urca processes are suppressed to different degrees, further complicating the calculation of the equilibrium condition.

\subsection{Equilibration via the strong interaction}

So far, we  considered the electron to be  the only negatively charged particle (muons were ignored for simplicity, but partake in similar reactions to the electrons), meaning that equilibration relied on the weak interaction. However, the situation may change as the temperature ramps up. For example, one may argue that we should account for the presence of thermal pions \cite{Fore2020}. A small population of pions could drastically change the story. At high temperatures neutrons may decay to protons and pions on the strong interaction timescale (which at $10^{-23}$~s is instantaneous for all practical purposes). The system would then reach an equilibrium where
\begin{equation}
   \mu_\mathrm{n}-\mu_\mathrm{p} = \mu_\pi
\end{equation}
with the pions subsequently equilibrating with the electrons on the weak interaction timescale (through pion decay). In the first estimates of this effect, Fore and Reddy~\cite{Fore2020} show that pion-equilibriated matter is more proton rich than the weak equilibria we have discussed. As a result, the equation of state may soften by as much at 10-15\%, which could have a significant impact on the merger dynamics. In addition, the presence of the thermal pions increases the heat capacity, leading to the matter cooling which may self-regulate the process. These are interesting ideas, but in order to test them with simulations we would need an equation of state that accounts for the thermal pions. In the absence of such a model, the best we can do is estimate if (and where) the thermal pions are likely to come into play. 

\begin{figure}[bt]
\centering
\includegraphics[width=0.95\textwidth]{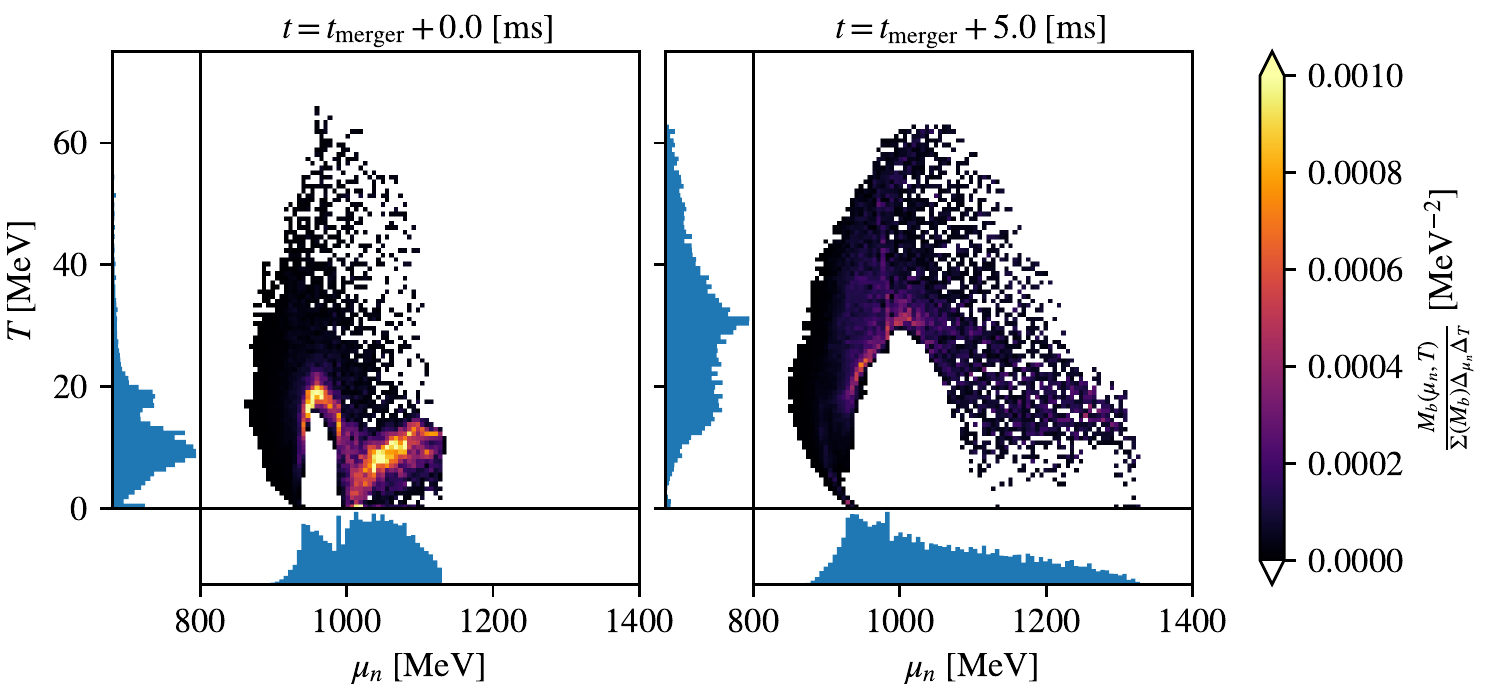}
\caption{\small Distribution of baryon mass $M_\mathrm{b}$ in the $\mu_\mathrm{n}$-$T$ plane for a simulation of the merger of two $1.4\,{M}_\odot$ neutron stars using the APR equation of state from \cite{Schneider2019}. The left panel coincides with the merger time, and the right panel is $5.0\,\mathrm{ms}$ post merger. The results suggest that  thermal pions, which may play a role at temperature above $T\gtrsim~ 25\,\mathrm{MeV}$~\cite{Fore2020}, may impact on a significant fraction of the simulated matter.}
\label{Fig:APRmu_nT}
\end{figure}

Drawing on the estimates from Fore and Reddy \cite{Fore2020}, which suggest that the effect becomes significant when $T\gtrsim~ 25\,\mathrm{MeV}$, we may consider the results in \cref{Fig:APRmu_nT}, which shows the distribution of matter in the $\mu_\mathrm{n}$-$T$ phase space, as indicative. We see that after merger the outflows from the core lead to a substantial migration of matter to higher chemical potentials, and the distribution of the matter has shifted to higher temperature, with most of the matter now above $T\gtrsim~ 15\,\mathrm{MeV}$ -- the spatial regions where this become important will be discussed in \cref{Sec:Summary}. Evidently, the temperature reaches the level where the pions may play a role throughout much of the matter. This suggests that we should seriously consider the role of the thermal pions, which inevitably involves moving towards a more complex matter model.

\section{Summary and outlook}
\label{Sec:Summary}

In order to better understand how the extreme physics of neutron star mergers is represented by large scale simulations, we have explored issues associated with finite temperature effects. While it is well-established that the matter equation of state softens when the temperate ramps up and that the thermal effects also regulate the extent to which the matter is transparent to neutrinos (with immediate impact on the ability of the neutrinos to remove energy from the system), the problem has more subtle aspects which have not (in our view) been considered at the level of detail they warrant. These aspects need attention if we aim towards a faithful representation of the physics, especially since they regulate the rates of nuclear reactions that enter discussion of bulk viscosity and neutrino leakage. Our particular focus was on the notion of  $\beta$-equilibrium in neutron star matter for the densities and temperatures reached in a neutron star merger. By  post-processing the results from an out-of-equilibrium merger simulation, for equal mass neutron stars described by the  APR equation of state \cite{Schneider2019}, we explored how different notions of equilibrium may affect the merger dynamics, and how this raises issues when attempting to account for the relevant conditions in a live simulation. We argued that this leads to a set of problems,  both computational and conceptual. In particular, our results suggest that the finite temperature effects lead to additional softening of the equation of state in some density regions, and to compositional changes that affect the processes that rely on deviation from equilibrium, such as bulk viscosity, both in terms of the magnitude and the equilibriation timescales. We have demonstrated that it is  far from straightforward to determine exactly which equilibrium conditions are relevant in which regions of the matter---the what and where of the problem---due to the dependence on neutrino absorption, further complicating the calculation of the reactions that work to restore the matter to equilibrium. 

\begin{figure}[bt]
\centering
\includegraphics[width=0.95\textwidth]{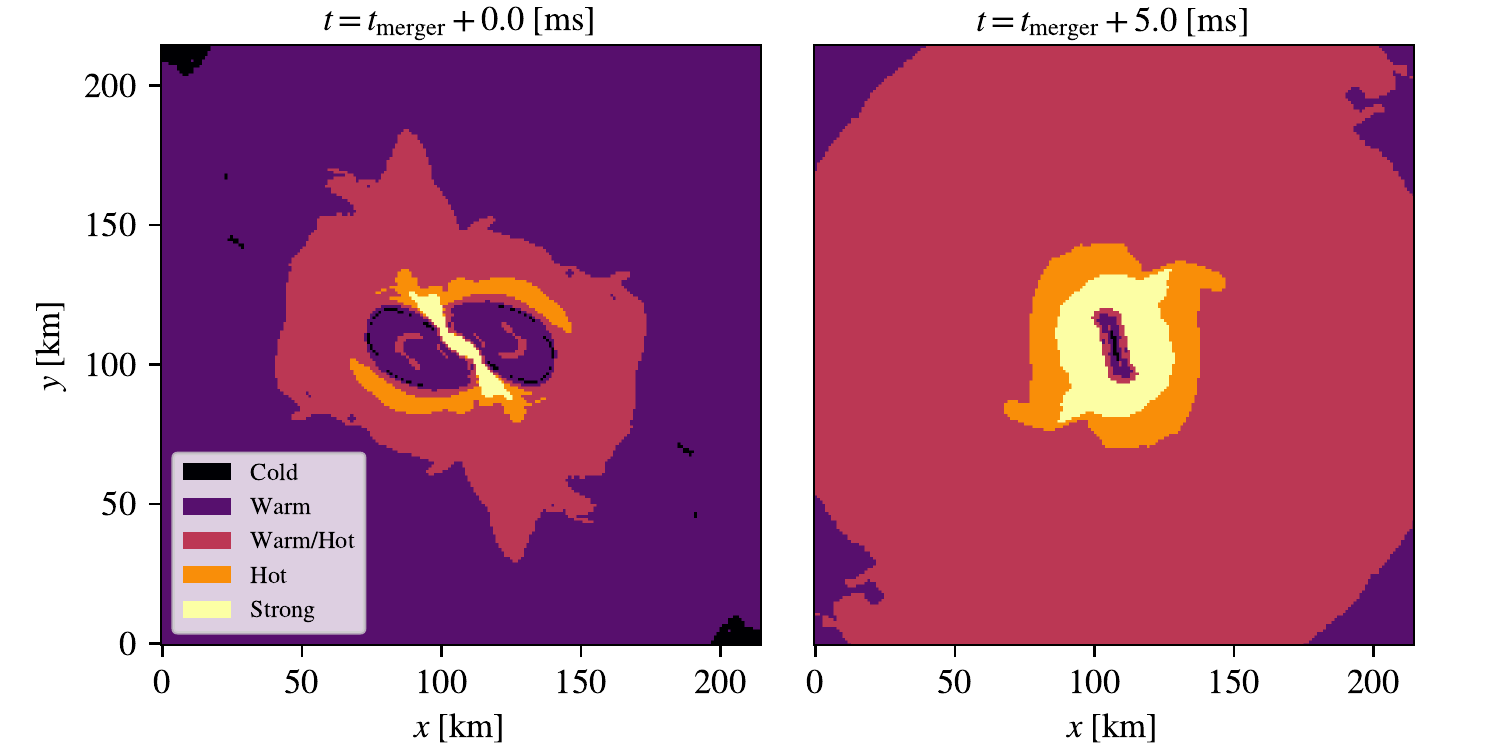}
\caption{Schematic of where the different equilibrium conditions discussed in \cref{Sec:Results} are expected to occur in a simulation of the merger of two $1.4\,{M}_\odot$ neutron stars using the APR equation of state from \cite{Schneider2019}. Left panel coincides with merger time, and right panel is $5.0\,\mathrm{ms}$ post-merger. The conditions on the matter are labelled as follows: ``Cold'' denotes matter with $T<1\,\mathrm{MeV}$ where the standard cold $\beta$ equilibrium condition  \cref{Eqn:ColdBetaEquilibrium} is relevant, ``Warm'' denotes matter with $T>1\,\mathrm{MeV}$ and $\tilde{\kappa}_\mathrm{ab} r_\mathrm{hot} < 1$ for both neutrino species where the warm $\beta$ equilibrium condition in \cref{Eqn:WarmBetaEquilibrium} is relevant, ``Hot'' denotes matter where $\tilde{\kappa}_\mathrm{ab} \Delta x > 1$ for both neutrino species, thus the hot $\beta$ equilibrium condition in \cref{Eqn:HotBetaEquilibrium} is relevant, ``Warm/Hot'' denotes matter with $T>1\,\mathrm{MeV}$ but with $\Delta x/r_\mathrm{hot}< \tilde{\kappa}_\mathrm{ab} \Delta x < 1$ for at least one of the neutrino species, where we cannot distinguish cleanly between hot and warm equilibrium, and ``Strong'' denotes matter with $T>25\,\mathrm{MeV}$ where pions may be relevant.}
\label{Fig:APRequilibria}
\end{figure}

As an attempt to summarize the discussion we provide the schematic illustration in \cref{Fig:APRequilibria}.  The figure identifies regions where the different equilibrium conditions discussed in \cref{Sec:Results} apply in our simulation (at merger and 5~ms later). We distinguish regions with ``cold''  matter ($T<1\,\mathrm{MeV}$) where the condition \eqref{Eqn:ColdBetaEquilibrium} applies, ``warm'' ($T>1\,\mathrm{MeV}$ and $\tilde{\kappa}_\mathrm{ab} r_\mathrm{hot} < 1$, the transparent regime, for both neutrino species) where we need to account for the imbalance of the nuclear reaction by adding $\mu_\delta$ as in \eqref{Eqn:WarmBetaEquilibrium}, a ``warm/hot''region which involves matter with $T>1\,\mathrm{MeV}$ but with $\Delta x/r_\mathrm{hot}< \tilde{\kappa}_\mathrm{ab} \Delta x < 1$ (neither locally trapped, nor transparent) for at least one of the neutrino species,  a ``hot'' region  where $\tilde{\kappa}_\mathrm{ab} \Delta x > 1$ (locally trapped) for both neutrino species, and finally a ``strong'' region  with matter at $T>25\,\mathrm{MeV}$ where one might anticipate thermal pions to play a role. As these regions evolve with time, a live identification of the appropriate local conditions is clearly far from trivial.

Admittedly, our discussion poses more problems than it solves. The main message is that we have to be careful when we consider how thermal effects enter the merger problem, especially if we aim toward a realistic implementation of the neutrinos with an on-the-fly calculation of the emission/absorption features. If the error bars associated with the inferred matter temperature are, indeed, as large as our results suggest then this issue is problematic. If we want to do better, the next steps are fairly clear. On the one hand we need to quantify the uncertainties associated with the temperature. While some of the aspects we are trying to model are likely sensitive to these errors, other are expected to be more robust. For example, the gravitational-wave signal may not be hugely affected by the issues we have discussed (although this remains to be quantified). The natural next step would be to work towards simulations that account for the different notions of equilibrium, either with a consistent model for $\mu_\delta$ for the chosen equation of state or with the thermal pions explicitly included. Such simulations are clearly within reach of our computational technology, but require some effort on the nuclear physics side. Similarly, we can take further steps towards a quantitative understanding of the relevance of bulk viscosity on the problem. A recent discussion of this effect \cite{2021arXiv210705094M} focused on the bulk viscous pressure (associated with the divergence of the fluid flow). Such results are indicative, but it is appropriate to raise a flag of caution. First of all, we need to keep in mind that the representation of the bulk viscosity as an effective contribution to the pressure is very much a perturbative notion. It is not at all clear that it remains appropriate for the nonlinear dynamics we are trying to model. Secondly, and perhaps more critically, the bulk viscosity arises as reactions strive to equilibrate the matter. Effectively, this leads to the different species fraction (like $Y_\mathrm{e}$ in our case) not being conserved. Now one has to be careful because the same effect is already accounted for in schemes aimed at modelling the neutrinos. Hence, adding a bulk viscous pressure may, in fact, lead to double counting. Clearly, these issues require further thinking.

\acknowledgements

Many colleagues have contributed useful discussion during this project. We would like to thank, in particular, Mark Alford, Constantinos Constantinou, Steven Harris, Madappa Prakash, Sanjay Reddy, Andreas Schmitt and Andre Schneider. We  are also grateful for support from STFC via grant numbers ST/R00045X/1 and ST/V000551/1. 

\bibliography{bib}

\newpage

\appendix
\section{Code Modifications}

\label{Sec:CodeModifications}
\subsection{Modifications to EOS\_Omni} \label{Sec:EOSOmniMod}

The EOS\_Omni module provides native support for a number of types of equation of state (including some tabulated), but there is no ability to use three-parameter tables in the format provided by the CompOSE database~\cite{CompOSE}. Therefore, we have written our own interface to tables in this form within the EOS\_Omni framework, and added the temperature inversion step required by some conservative-to-primitive methods.

We use a set of tricubic interpolators (one for each tabulated variable,  $p$, $\epsilon$, and $c^2_s$) defined piece-wise for each cell in the table grid. In order to account for the large range of values of the table parameters $\rho$ and $T$, and all of the tabulated variables we use the logarithm instead or the raw value. An issue arises with $\epsilon$ as, unlike $p$ and $c^2_s$, it is not everywhere positive. Hence, instead of using $\log\left(\epsilon\right)$ we use $\mathrm{arcsinh}\left(\epsilon\right)$, which behaves like $\log$ for $\left|\epsilon\right|\gg 1$, but is also defined for $\epsilon \leq 0$. 

The interpolators each take the form (for a  general variable $f$ to be interpolated)
\begin{align}
    f\left( \log\rho, \log T, Y_\mathrm{e} \right) &= \sum^3_{i=0} \sum^3_{j=0} \sum^3_{k=0} a_{ijk} {x}^i {y}^j {z}^k, \label{Eqn:InterpolatorForm}
\end{align}
where $\left(x, y, z\right)$ are the shifted and normalised values of $\left( \log\rho, \log T, Y_\mathrm{e} \right)$ with respect to the table cell that contains $\left( \log\rho, \log T, Y_\mathrm{e} \right)$ scaled such that each cell runs from $0\rightarrow 1$ in all three dimensions. The polynomial coefficients $a_{ijk}$ are calculated using the values of the variable $f$ and all derivatives that contain at most first order terms in each of the three parameters, evaluated at each of the eight corners of the grid cell. These $64$ values are arranged in a vector $\boldsymbol{m}$ which we relate to a vector containing the elements $a_{ijk}$, $\boldsymbol{a}$, through a matrix $\boldsymbol{B}$, giving
\begin{align}
    \boldsymbol{m} &= \boldsymbol{B} \boldsymbol{a} \label{Eqn:InterpolatorCoeffs},
\end{align}
where we now need to find the elements of $\boldsymbol{B}$. This is done by applying the same derivative terms in $\boldsymbol{m}$ to \cref{Eqn:InterpolatorForm} and reading off the numerical coefficients that appear in front of each of the elements of $\boldsymbol{a}$ for each of the elements of $\boldsymbol{m}$. The final step is to invert \cref{Eqn:InterpolatorCoeffs} to give 
\begin{align}
    \boldsymbol{a} &= \boldsymbol{B}^{-1} \boldsymbol{m},
\end{align}
where the interpolation coefficients are now defined in terms of the table variable $f$ and its derivatives with respect to the other table parameters. Constructing the table in this form also allows easy calculation of derivatives of the table variables with respect to the table values, however care must be taken to undo the transformations (taking the $\log$ or $\mathrm{arcsinh}$) that were made when creating the table in the first place.

The temperature inversion step is required when the known values are $\left(\rho,\epsilon,Y_\mathrm{e}\right)$, and the value of $T$ must be calculated. This is used by one of the conservative-to-primitive methods we mention below, in particular the more stable backup method. As this step is only required in the backup method, we do not need to ensure the inversion step is fast, it is significantly more important that it be robust. Hence we use an exhaustive search over all possible values of $T$ (in this context meaning within the limits of the table). 

As the values of $\rho$ and $Y_\mathrm{e}$ are known, we can evaluate those two respective sums in \cref{Eqn:InterpolatorForm} to leave a cubic polynomial in $\log T$ for each cell in the table. We then use a standard cubic polynomial solver to find which of these have solutions to $\epsilon = \epsilon\left(\rho,T,Y_\mathrm{e}\right)$ in the correct limits by disregarding solutions for a given cell that lie outside said cell, which leaves us with the value of $T$ that corresponds to the input $\epsilon$ given $\rho$ and $Y_\mathrm{e}$. There exists the possibility for multiple solutions given an arbitrary equation of state table, so if this does occur we choose $T$ to minimise the change in $T$ compared to the value on the simulation grid at the previous timestep.

\subsection{Modifications to GRHydro} \label{Sec:GRHydroMod}

Modern GRMHD codes utilise the \textit{Valencia formulation}~\cite{Banyuls1997, Baiotti2005} of the relativistic hydrodynamics equations for the evolution of the fluid, which take the form (neglecting the magnetic field component)
\begin{align}
\partial_t \left( \sqrt{\gamma} \boldsymbol{q} \right) + \partial_i \left( \alpha \sqrt{\gamma} \boldsymbol{f}^i \left( \boldsymbol{p} \right) \right) = \boldsymbol{s} \left( \boldsymbol{p} \right),
\end{align}
where $\alpha$ and $\gamma$ are taken from the 4-metric, $\boldsymbol{q}$ denotes a vector of \textit{conserved variables}
\begin{align}
\boldsymbol{q} &= \left( D, S_i, \tau, DY_\mathrm{e} \right),
\end{align}
and $\boldsymbol{p}$ denotes a vector of \textit{primitive variables}
\begin{align}
\boldsymbol{p} &= \left( \rho, v^i, T, Y_\mathrm{e} \right).
\end{align}

The conserved variables are given analytically in terms of the elements of $\boldsymbol{p}$ through
\begin{align}
\boldsymbol{q}\left( \boldsymbol{p} \right) = 
\begin{pmatrix} 
D \\
S_i \\
\tau \\
DY_\mathrm{e}
\end{pmatrix} = 
\begin{pmatrix} 
\rho W \\
\rho h W^2 v_i \\
\rho h W^2 - p - \rho W \\
\rho W Y_\mathrm{e}
\end{pmatrix}, \label{Eqn:Prim2Con}  
\end{align}
where the isotropic pressure $p$ and specific enthalpy density $h=1+\epsilon+p/\rho$ are calculated from the equation of state, and the Lorentz factor $W = (1 - v^i v_i)^{-1/2}$. As the evolution of $\boldsymbol{q}$ depends on $\boldsymbol{p}$ we must calculate $\boldsymbol{q}\left( \boldsymbol{p} \right)$ at each step. As this is not possible analytically, a number of numerical schemes have been developed (see Siegel et al.~\cite{Siegel2018} for a recent summary).

The conservative-to-primitive inversion method used by GRHydro relies on the availability of the derivatives $\partial p/\partial \rho$ and $\partial p/\partial \epsilon$ from the EoS. However, most three-parameter EoSs are tabulated by $\rho$, $T$, and $Y_\mathrm{e}$, with $\epsilon$ a function of these parameters. This means that $\partial p/\partial \epsilon$ must be calculated through
\begin{align}
\frac{\partial p}{\partial \epsilon} &= \frac{\partial p}{\partial T}\frac{\partial T}{\partial \epsilon} + \frac{\partial p}{\partial \rho}\frac{\partial \rho}{\partial \epsilon} + \frac{\partial p}{\partial Y_\mathrm{e}}\frac{\partial Y_\mathrm{e}}{\partial \epsilon} \nonumber \\
&= \frac{\partial p}{\partial T}\left(\frac{\partial \epsilon}{\partial T}\right)^{-1} + \frac{\partial p}{\partial \rho}\left(\frac{\partial \epsilon}{\partial \rho}\right)^{-1} + \frac{\partial p}{\partial Y_\mathrm{e}}\left(\frac{\partial \epsilon}{\partial Y_\mathrm{e}}\right)^{-1}
\end{align}
which, when any of the $\partial \epsilon /\partial X$ terms approaches $0$, can explode to infinity, leading to instability.

To improve upon this, we have implemented two methods from Siegel et al.~\cite{Siegel2018} which are formulated in terms more suited for $\left(\rho,T,Y_\mathrm{e}\right)$ tables. The first is based on the 2-dimensional Newton-Raphson scheme presented by Anton et al.~\cite{Anton2006}, and solves for $W$ and $T$. The second method was suggested by Palenzuela et al.~\cite{Palenzuela2015} and solves for $x=hW$ on a bracketed interval, where we use Dekker's method for the bracketed root finding. These two methods were chosen for a balance of speed and stability. The faster 2-D Newton-Raphson scheme is attempted first, and if it fails then the more stable 1-D Dekker method is used. 

It is possible that the evolution of the conserved variables will cause them to no longer correspond to a physical solution within the bounds of the equation of state table. This typically occurs when the effective temperature of the fluid falls below the lowest temperature available in the equation of state table $T_{\mathrm{min}}$. In this case, we assume $T = T_{\mathrm{min}}$ and solve for $W$. Due to this mismatch in temperatures, we know that the primitives we obtain will not correspond exactly to the conservatives with which we started, and so we must decide where we want this error to lie. 

We experimented with trying to minimise the total relative error across the conservatives, but found this introduced spurious kinetic energy into the simulation. Instead, we do our best to conserve the $3-$momentum density $p^i = \rho W v^i = D v^i$. As the value of $W$ will not impact on the direction of $p^i$, we can instead choose to attempt to minimise the error in $p^2 = p_i p^i = D^2 v^2$, and because $D$ does not depend on the equation of state we can safely ignore it. Therefore, we must solve to minimise error in $v^2$. 

The combination of conserved variables closest to $v^2$ is
\begin{equation}
    \frac{S^2}{\left( \tau + D \right)^2} = \frac{z^2 v^2}{\left( z - p \right)^2} = v^2 \frac{1}{\left( 1 - \frac{p}{z} \right)^2} \label{Backup_LHS},
\end{equation} 
where $z=\rho h W^2$, which reduces to $v^2$ for $p \ll z$. We can also calculate this quantity from the primitive variables using
\begin{align}
    v^2 \frac{1}{\left( 1 - \frac{p}{z} \right)^2} &= \frac{W^2-1}{W^2 \left( 1 - \frac{p(W)}{z(W)} \right)^2}, \label{Backup_RHS}
\end{align}
where $p(W)$ and $z(W)$ are calculated from the EoS using $\left(\rho=D/W,T=T_\mathrm{min},Y_\mathrm{e} = DY_\mathrm{e} / D\right)$. Combining \cref{Backup_LHS,Backup_RHS} we define an error function $f$, which measures the difference between the input $S^2/\left(\tau + D\right)^2$ and the value calculated from our $W$ guess, which takes the form
\begin{align}
    f(W) &= \log \left( \frac{W^2-1}{W^2 \left( 1 - \frac{p(W)}{z(W)} \right)^2} \right) - \log \left( \frac{S^2}{\left( \tau + D \right)^2} \right),
\end{align}
where we have taken the log of each equation in order to improve behaviour far from $f(W) = 0$. We also obtain better behaviour where $v^2 \ll 1 \Rightarrow W \approx 1$ by using $w = \log(W - 1)$, which yields
\begin{align}
    f(w) &= \log \left( \frac{e^{2w} + 2e^{w}}{\left( e^{2w} + 2e^w + 1 \right) \left( 1 - \frac{p(w)}{z(w)} \right)^2} \right) - \log \left( \frac{S^2}{\left( \tau + D \right)^2} \right).
\end{align}

$W$ is physically constrained by the lower bound $1\leq W$, which corresponds to $-\infty \leq w$. However the limits of floating point number representation limit meaningful values of $W>1$ to 
\begin{align}
    W &= 1 + \varepsilon,
\end{align} where $\varepsilon$ is the machine epsilon (in the case of double precision floating point numbers, $\varepsilon = 2^{-52} \approx 2.22\times 10^{-16}$), thus giving a lower bound for $w$ of 
\begin{align}
    w_\mathrm{min} &= \log{\varepsilon}.
\end{align} 
We can obtain an upper bound for $w$ by considering the fact that if $\rho$ drops below an atmosphere density $\rho_\mathrm{atmo}$, the simulation will consider that point to be part of the atmosphere and reset all of the fluid variables to pre-defined defaults. This means that for the point in question $w$ must be such that 
\begin{align}
    \frac{D}{\left( 1+e^w \right)} &\geq \rho_\mathrm{atmo},
\end{align}
to avoid resetting, which we can rearrange to 
\begin{align}
    w_\mathrm{max} = \log \left(\frac{D}{\rho_\mathrm{atmo}} - 1 \right).
\end{align}

Using the two bounds for $w$ and the function $f(w)$, we again use Dekker's method to solve for $w$ such that $f(w)=0$. If we find there is no solution for $f(w)=0$ on the interval $w_\mathrm{min} \leq w \leq w_\mathrm{max}$, we use a modified objective function $g(w)$, defined by
\begin{align}
     g(w) &= \left( \frac{e^{2w} + 2e^{w}}{\left( e^{2w} + 2e^w + 1 \right) \left( 1 - \frac{p(w)}{z(w)} \right)^2} \right) - \left( \frac{S^2}{\left( \tau + D \right)^2} \right),
\end{align}
to determine whether there is a solution on the interval $-\infty \leq w \leq w_\mathrm{min}$ by observing that 
\begin{align}
    \lim_{w\to -\infty} g(w) &= \left( \frac{0}{0 + 1} \frac{1}{\left( 1 - \frac{p(w)}{z(w)} \right)^2} \right) - \frac{S^2}{\left( \tau + D \right)^2} \nonumber \\
    &= - \frac{S^2}{\left( \tau + D \right)^2}
\end{align} 
If we find that $g(w_\mathrm{min})\times g(-\infty) \leq 0$, then we accept $W=1$ as the solution. We can also test for a solution on the interval $w_\mathrm{max} \leq w \leq \infty$ by observing that 
\begin{align}
    \lim_{w\to\infty} g\left(\infty\right) &= \lim_{w\to\infty}\left( \frac{e^{2w} + 2e^{w}}{\left( e^{2w} + 2e^w + 1 \right)} \times \left(1 -  \frac{1}{W^2} \frac{1}{\frac{\rho \left( 1 + \epsilon \right)}{p} + 1} \right)^{-2} \right) - \left( \frac{S^2}{\left( \tau + D \right)^2} \right),
\end{align}
where we can immediately see that
\begin{align}
    \lim_{w\to\infty}\left( \frac{e^{2w} + 2e^{w}}{\left( e^{2w} + 2e^w + 1 \right)} \right) &= 1,
\end{align}
and we can use the dominant energy condition
\begin{equation}
p^2 \leq \rho^2 \left(1 + \epsilon \right)^2,
\end{equation}
to show that
\begin{align}
    \frac{1}{\frac{\rho \left( 1 + \epsilon \right)}{p} + 1} &\leq \frac{1}{2}.
\end{align}
This leaves us with
\begin{align}
    \lim_{w\to\infty} g(w) &= \left( 1\times \left( 1 - \lim\limits_{w\to\infty} \frac{1}{2 \left(1+e^w\right)^2} \right)^{-2} \right) - \left( \frac{S^2}{\left( \tau + D \right)^2} \right) \\
    &= 1 - \left( \frac{S^2}{\left( \tau + D \right)^2} \right).
\end{align}
and so if $g(w_\mathrm{max}) \times g(\infty) \leq 0$ then the solution is sufficiently large that the point will be set to atmosphere, and this is accepted as a solution. If both of these tests fail, then Con2Prim fails, and the simulation is aborted.

Once we have a solution for $W$, we can use this and $T_\mathrm{min}$ to calculate the values of the primitives through
\begin{align}
    \rho &= \frac{D}{W}, \\
    Y_\mathrm{e} &= \frac{DYe}{D}, \\
    p &= p \left(\rho,T_\mathrm{min},Y_\mathrm{e}\right), \\
    v^i &= \frac{\gamma^{ij} S_j}{\tau + D + p}\ .
\end{align}
As we know the primitive variables obtained will be inconsistent with the conservative variables with which we started, we calculate the values of the conserved variables using \cref{Eqn:Prim2Con}, and use these to replace the previous values.
\end{document}